\documentstyle[aps,multicol,eqsecnum,epsf]{revtex}
\def\dir{.}
\begin{document}
\draft
\title{Anisotropic Scaling in Threshold Critical Dynamics of \\ 
Driven Directed Lines}
\author{Deniz Erta\c s\footnote{Present Address: Lyman Laboratory of 
Physics, Harvard University, Cambridge, MA 02138} and Mehran Kardar}
\address{Department of Physics\\
Massachusetts Institute of Technology\\
Cambridge, Massachusetts 02139}
\date{\today}
\maketitle
\begin{abstract}
The dynamical critical behavior of a single directed line 
driven in a random medium near the depinning threshold 
is studied both analytically (by 
renormalization group) and numerically, in the context of a Flux Line 
in a Type-II superconductor with a bulk current 
$\vec J$. In the absence of transverse fluctuations, the system reduces
to recently studied models of interface depinning. 
In most cases, the presence of transverse fluctuations are
found not to influence the critical exponents that describe 
longitudinal correlations.   
For a manifold with $d=4-\epsilon$ internal dimensions,
longitudinal fluctuations {\it in an isotropic medium}
are described by a roughness exponent 
$\zeta_\parallel=\epsilon/3$ to all orders in $\epsilon$, and
a dynamical exponent $z_\parallel=2-2\epsilon/9+O(\epsilon^2)$. 
Transverse fluctuations have a distinct and smaller roughness exponent
$\zeta_\perp=\zeta_\parallel-d/2$ for an isotropic medium. 
Furthermore, their relaxation is  much  slower, characterized
by a dynamical exponent $z_\perp=z_\parallel+1/\nu$, where 
$\nu=1/(2-\zeta_\parallel)$ is the correlation length exponent.
The predicted exponents agree well with 
numerical results for a flux line in three dimensions. 
As in the case of interface depinning models, anisotropy leads to
additional universality classes. A nonzero Hall angle, 
which has no analogue in the interface models, also affects the 
critical behavior.
\end{abstract}

\begin{multicols}{2}

\section{Introduction and Summary}

The study of dynamical critical phenomena associated with the 
pinning-depinning transition in random media has become a subject of 
considerable interest in recent years. This is due to the importance 
of pinning in a wide variety of technologically important phenomena 
such as flux line (FL) motion in Type-II superconductors, dynamics of
interfaces (phase boundaries, invasion fronts, cracks, surface growth,
to name a few), and charge-density wave (CDW) transport.
These systems are characterized by a rough energy landscape due to the 
randomness in the medium. At zero temperature there are 
two distinct ``phases", distinguished by an order parameter
(henceforth called velocity) that measures the dynamic response, such 
as the average velocity for a FL, or current for a CDW. For small 
driving forces, the system is trapped by one of the many available 
metastable stationary states, and is ``pinned" to the 
impurities in the medium. Critical behavior emerges as the 
stationary states disappear, and the system starts moving with a 
nonzero velocity, when the driving force is increased above a 
threshold value. Extensive experimental
\cite{Robbins}, theoretical\cite{Fisher85,NF92,Middleton}, and 
simulation\cite{Myers} work has been done to understand the 
properties of this transition in  CDW systems.
There are also numerous studies on the depinning of driven
interfaces\cite{Bruinsma,NSTL,NF93,Martys,Buldyrev,Leschhorn}.  
A better theoretical understanding of this dynamical
phase transition was recently achieved, and critical exponents
were calculated through an $\epsilon$-expansion for both CDW 
systems\cite{NF92} and driven interfaces\cite{NSTL,NF93}. 
More recently, we performed similar calculations for the depinning of 
an elastic line in a bulk random medium, like a polymer in a gel network, 
a FL in a type-II superconductor, or a screw dislocation in a 
crystal\cite{EKfluxdepin}. 
In this article, we present a detailed report of our
study on the dynamical critical behavior associated with the depinning 
of a FL, and in general on the depinning of directed 
manifolds in random media, through methods similar to those used for 
CDWs and interfaces.

\begin{figure}
\narrowtext
\centerline{\epsfxsize=2.9truein
\epsffile{\dir/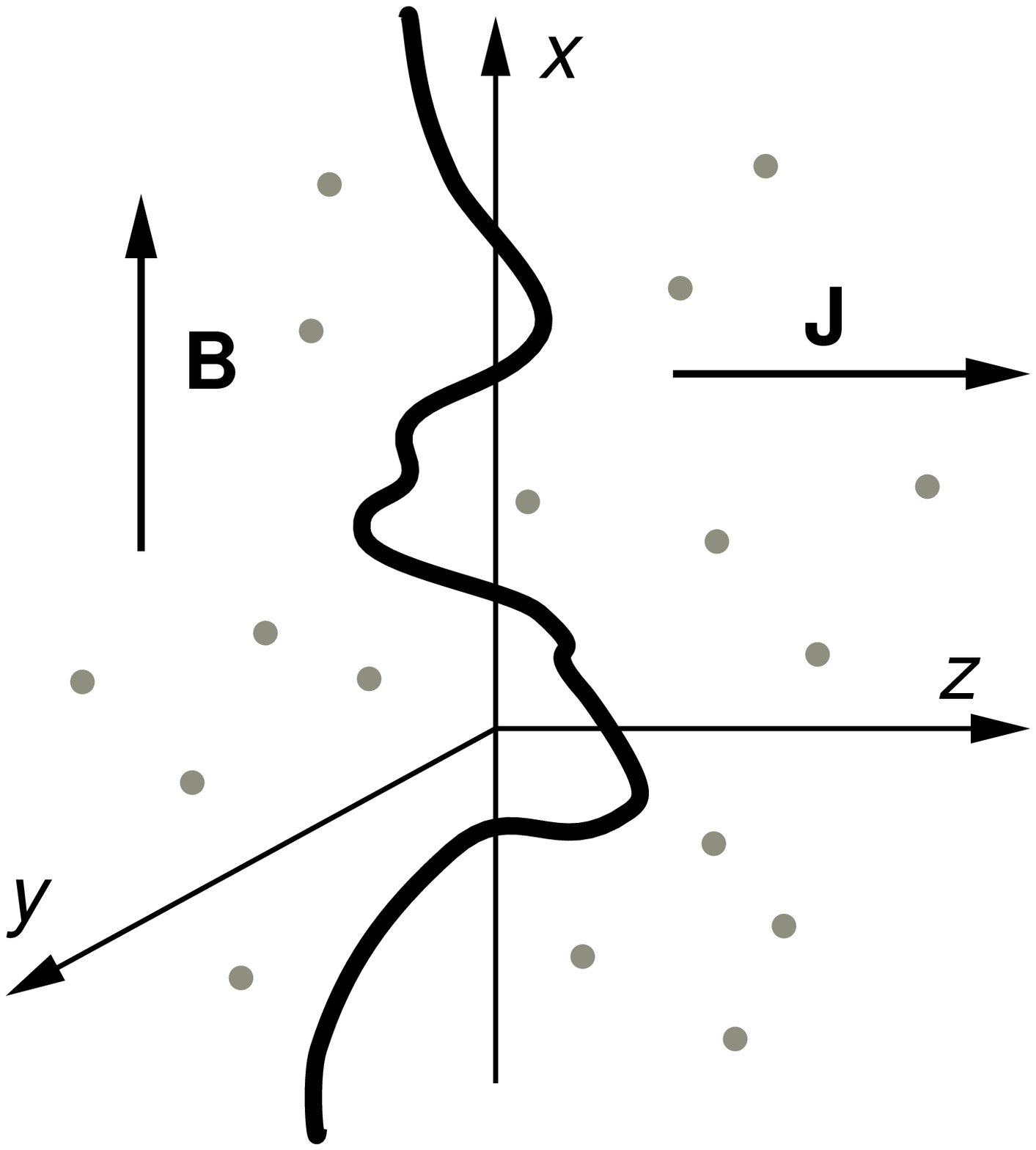}}
\centerline{}
\centerline{\bf (a)}
\centerline{}
\centerline{\epsfxsize=2.9truein
\epsffile{\dir/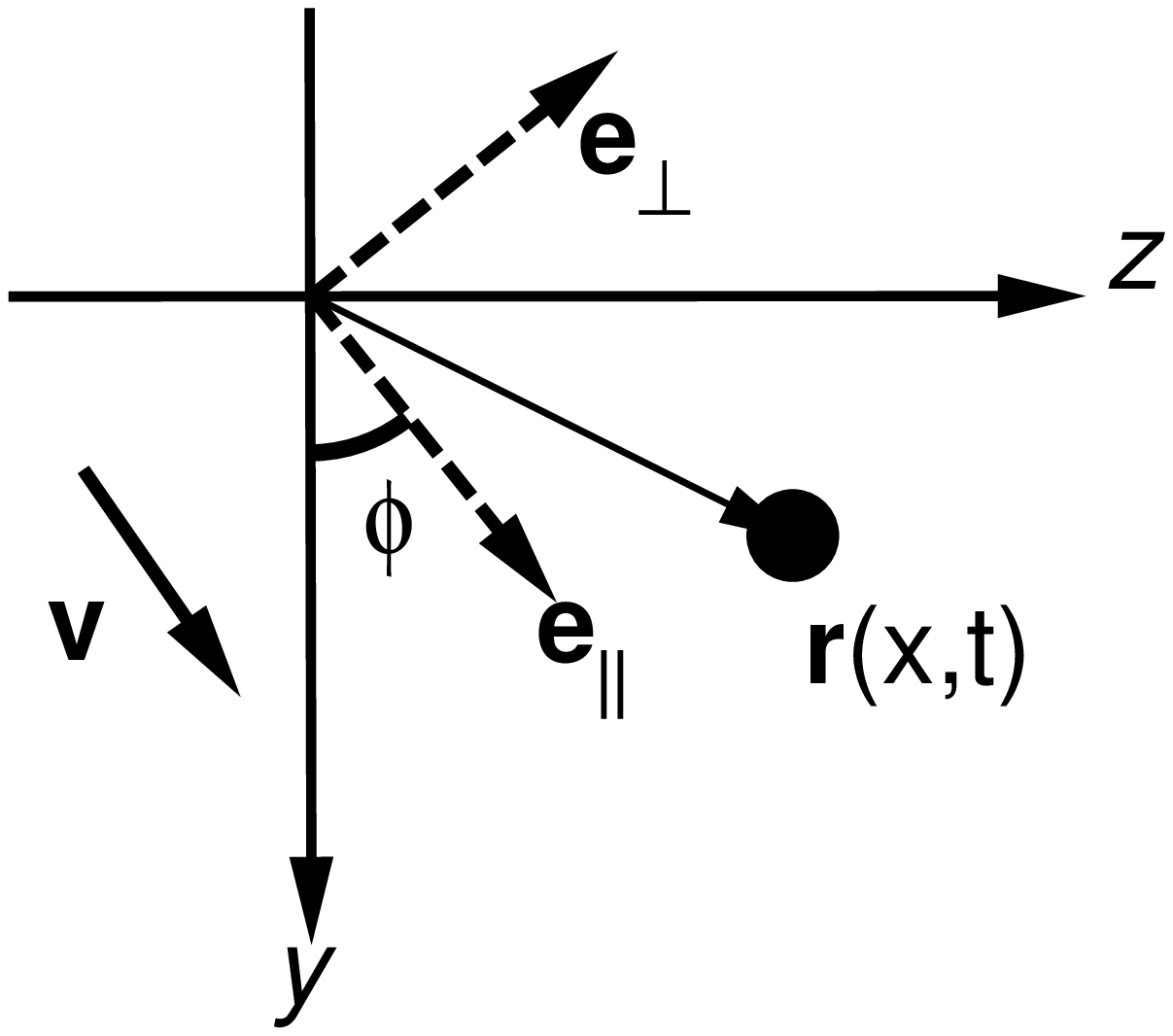}}
\centerline{}
\centerline{\bf (b)}
\centerline{}
\caption{Geometry of the FL in a medium with impurities: 
(a) Three-dimensional geometry. (b) A cross section of the 
medium at fixed $x$. The average drift velocity ${\bf v}=v{\bf e_\parallel}$ 
makes an angle $\phi$ with the $y-$axis.}
\label{geometry}
\end{figure}

Specifically, let us consider the geometry of the FL shown in 
Fig.~\ref{geometry}. The superconductor is subject to a magnetic
field $\vec B=B{\hat x}$ along the $x$-axis, and a bulk supercurrent 
$\vec J=J{\hat z}$ along the $z$-axis. A FL is oriented along $\vec B$ 
on the average, but deviates from a straight line due to impurities 
in the superconductor, which are represented by a potential 
$V(x,y,z)$. The conformations of the FL are described
by $\vec R(x,t)=x{\hat x}+{\bf r}(x,t)$, where 
${\bf r}(x,t)=y(x,t){\hat y}+z(x,t){\hat z}$ is a {\it two component}
vector, lying in a plane normal to the magnetic field. 
The bulk current $\vec J$ drives the FL along the $y$-direction
through the Lorentz Force $\vec F_L = \Phi_0 \vec J\times \vec B$. 
($\Phi_0$ is the flux quantum.) If the bulk current is large
enough, the FL drifts with an average velocity ${\bf v}$. Due to the chiral 
nature of the supercurrents around the FL, ${\bf v}$ is in general not
along the $y$-direction, but makes an angle $\phi$ with the $y$-axis.
This is usually called the {\it Hall angle}, and although 
typically small\cite{Graybeal}, it can be significant near the depinning 
transition.

It is more convenient to work with components of ${\bf r}$ that are parallel 
and perpendicular to ${\bf v}$, i.e.
\begin{equation}
{\bf r}(x,t)={r_\parallel}(x,t){\bf e_\parallel}+{r_\perp}(x,t){\bf e_\perp},
\end{equation}

\noindent where the unit vectors ${\bf e_\parallel}$ and ${\bf e_\perp}$ 
are rotated by $\phi$ 
from the $y$- and $z$-axes respectively, as shown in Fig.~\ref{geometry}b.
In Sec.\ref{eom} we show that, under very general
assumptions, the equation of motion for small deviations
around a straight line, generalized to $d$-dimesional internal 
coordinates ${\bf x}\in\Re^d$, can be written as
\begin{mathletters}
\label{motion}
\begin{eqnarray}
\eta\partial_t{r_\parallel}\!&=&\!K_{11}\nabla_{\bf x}^2{r_\parallel}\!
+K_{12}\nabla_{\bf x}^2{r_\perp}\!+F
\!+\!\tilde {f_\parallel}({\bf x},{\bf r}({\bf x},t)), \\
\eta\partial_t{r_\perp}\!&=&\!K_{21}\nabla_{\bf x}^2{r_\parallel}\!
+K_{22}\nabla_{\bf x}^2{r_\perp}
\!+\!\tilde {f_\perp}({\bf x},{\bf r}({\bf x},t)),
\end{eqnarray}
\end{mathletters}
where $\eta$ is the viscosity the FL and $F=\Phi_0J$. The moduli
$K_{\alpha\gamma}$ relate the elastic force to the local curvature and are 
in general nondiagonal for a sample with orientation-dependent
core energy, or nonzero Hall angle (cf. Sec.~\ref{eom}). 
The random forces $\tilde f_\alpha$ that arise from the
impurity potential $V$ are taken to have zero mean with correlations
\begin{equation}
\label{statistics}
\langle \tilde f_\alpha({\bf x},{\bf r})\tilde f_\gamma({\bf x}',{\bf r}')
\rangle=
\delta^d({\bf x}-{\bf x}')\tilde\Delta_{\alpha\gamma}({\bf r}-{\bf r}'),
\end{equation}
where $\tilde\Delta$ is a function that decays rapidly for large 
values of its argument. (The indices $\alpha,\gamma,\dots=
\{\parallel,\perp\}$.) 

Ignoring fluctuations of the FL transverse to the direction of
average velocity, i.e. setting ${r_\perp}=0$, leads to an interface
depinning model  studied by Nattermann, Stepanow, Tang, and
Leschhorn (NSTL)\cite{NSTL}, and by Narayan and Fisher
(NF)\cite{NF93}. Hence, the major difference between Eqs.(\ref{motion}) 
(henceforth called the ``vector depinning model") and the previously 
studied ``interface model" is the existence of transverse fluctuations,
making the position of the line ${\bf r}$ a vector instead of a 
scalar ``height" variable. The effects of such transverse fluctuations 
for large driving forces and average velocities, when the randomness 
in the medium can be approximated as uncorrelated in space and time,
were shown\cite{EKlines,EKpoly} to create a much richer dynamical phase 
diagram than the corresponding interface growth model, namely the
Kardar-Parisi-Zhang (KPZ) equation\cite{KPZ}. Then, the natural questions 
to ask are: How do these transverse fluctuations scale near the depinning 
threshold, and how do they influence the critical dynamics of 
longitudinal fluctuations?

In order to make these questions more quantifiable, we consider the 
exponents that characterize the critical behavior near the depinning
transition. Let ${\bf F}({\bf v})$ denote the driving force required to move 
the FL with a velocity ${\bf v}=v{\bf e_\parallel}$. For small values of 
$F=|{\bf F}|$, 
the line is pinned by the disorder in the medium. There is a threshold force 
$F_c$, such that the line moves with a nonzero average velocity 
$v$ iff $F>F_c$\cite{fcfoot}. For $F$ slightly above threshold, we expect 
the average velocity to scale as
\begin{equation}
v=A(F-F_c)^\beta,
\end{equation}
where $\beta$ is the velocity exponent and $A$ is a nonuniversal
constant. Superposed on the steady advance of the line 
are rapid ``jumps" as portions of the line depin from strong pinning 
centers. Such jumps are similar to avalanches in other slowly forced 
systems and have a power-law distribution in size, cut off at a 
characteristic correlation length $\xi$. On approaching the threshold,
$\xi$ diverges as
\begin{equation}
\xi\sim (F-F_c)^{-\nu},
\end{equation}
defining a correlation length exponent $\nu$. At length scales
up to $\xi$, the interface is self-affine,
with correlations satisfying the dynamic scaling form
\begin{eqnarray}
\langle[{r_\parallel}({\bf x},t)-{r_\parallel}({\bf 0},0)]^2\rangle 
&=& |{\bf x}|^{2\zeta_\parallel}g_\parallel(t/|{\bf x}|^{z_\parallel}),\\
\langle[{r_\perp}({\bf x},t)-{r_\perp}({\bf 0},0)]^2\rangle 
&=& |{\bf x}|^{2\zeta_\perp}g_\perp(t/|{\bf x}|^{z_\perp}), 
\end{eqnarray}
where $\zeta_\alpha$ and $z_\alpha$ are roughness and 
dynamic exponents, respectively. The scaling functions 
$g_\alpha$ go to a constant as their arguments approach 0;
$\zeta_\parallel$ and $\zeta_\perp$ are the longitudinal and
transverse wandering exponents of an instantaneus line profile;
$z_\parallel$ and $z_\perp$ characterize scaling of relaxation times of 
longitudinal and transverse modes with wave vector
through $\tau_\alpha(q)\sim q^{-z_\alpha}$.
Beyond the length scale $\xi$, regions move more or less
independently of each other and the system is no longer critical.
The behavior of the moving line is described by the exponents
calculated earlier\cite{EKlines,EKpoly} for time dependent noise.
Ignoring any potential nonlinearities leads to a regular
diffusion equation with white noise, for which the roughness and dynamic 
exponents are  $\zeta^+_\parallel=\zeta^+_\perp=(2-d)/2,\,z^+=2$.
In the interface model, transverse fluctuations do not exist,
thus, $\zeta_\perp$ and $z_\perp$ are not defined.

Equations (\ref{motion}) can be analyzed using the formalism of
Martin, Siggia, and Rose (MSR)\cite{MSR}. A renormalization group (RG) 
treatment of the ``interface model", studied by NSTL\cite{NSTL} and
NF\cite{NF93}, indicates an upper critical dimension of $d_c=4$, and 
exponents in $d=4-\epsilon$ dimensions, given to one-loop order as
$\zeta=\epsilon/3$ and $z=2-2\epsilon/9$.
NSTL obtained this result by directly averaging the MSR
generating functional $Z$, and calculating the renormalization
of the force-force correlation function $\tilde\Delta(r)$.
NF, on the other hand, expanded $Z$ around a saddle point solution
corresponding to a mean-field approximation\cite{SZ} to
Eqs.(\ref{motion}) which involves {\it temporal}
force-force correlations $C(vt)$. They point out some of the
deficiencies of conventional low-frequency analysis, and 
suggest that the roughness exponent is equal to 
$\epsilon/3$ to all orders in perturbation theory.
They also show that for two different classes of disordered systems,
random-field  and random-bond disorder, the zero temperature
interface dynamics is essentially the same near threshold.
Their argument remains valid for vector depinning, and our results 
will be applicable to both types of randomness. As we shall 
demonstrate in Section~\ref{model}, the longitudinal exponents of the
``vector" model are identical to those of the depinning interface,
and given by
\begin{eqnarray}
\label{exp1}
\zeta_\parallel&=&\epsilon/3,\\
\label{exp2}
z_\parallel&=&2-2\epsilon/9+O(\epsilon^2).
\end{eqnarray}
Other exponents are determined by exact exponent identities from 
$\zeta_\parallel$ and $z_\parallel$ as
\begin{eqnarray}
\label{exp3}
\nu&=&\frac{1}{2-\zeta_\parallel}=\frac{3}{6-\epsilon}, \\
\label{exp4}
\beta&=&(z_\parallel-\zeta_\parallel)\nu=1-\epsilon/9+O(\epsilon^2).
\end{eqnarray}

Following the formalism of NF, we employ a perturbative expansion 
of the disorder-averaged MSR partition function around a mean-field
solution for scalloped impurity potentials\cite{NF93}. We show that 
slightly above threshold, transverse fluctuations do not 
significantly affect the dynamics of longitudinal fluctuations, apart
from shifting the threshold force $F_c$. Specifically, the exponents
and exponent identities given in Eqs.~(\ref{exp1}--\ref{exp4}) for 
$d<d_c$ are also correct for the vector depinning model. However, 
transverse fluctuations turn out to scale differently, with 
$\zeta_\perp\neq\zeta_\parallel$ and $z_\perp\neq z_\parallel$. 
In particular, in an {\it isotropic} medium with Hall angle $\phi=0$
(Model A in Section~\ref{eom}), the renormalization of
transverse temporal force-force correlations $C_\perp(vt)$ yields 
\begin{equation}
\label{exp5}
\zeta_\perp=\zeta_\parallel-\frac{d}{2}=-2+\frac{5\epsilon}{6}, \\
\end{equation}
correct to all orders in $\epsilon$. 
The transverse dynamic exponent is given by an {\it exact} exponent 
identity:
\begin{equation}
\label{exp6}
z_\perp=z_\parallel+\frac{1}{\nu}=4-\frac{5\epsilon}{9}+O(\epsilon^2).
\end{equation}
These conclusions can also be generalized to  
more than one transverse direction: the results do not depend on 
the number of transverse coordinates.
For the FL $(\epsilon=3)$, the critical exponents are then predicted to be
\begin{equation}
\begin{array}{lll}
\zeta_\parallel=1,&z_\parallel\approx4/3,&\nu=1, \\
\beta\approx1/3,&\zeta_\perp=1/2,&z_\perp\approx7/3.
\end{array}
\end{equation}
This implies that in 
a type II superconductor driven slightly above threshold,
flux lines are contained mostly in the plane normal to the current, up
to the correlation length scale $\xi$. This may have a noticeable
effect on the dynamics of entanglement of flux lines near depinning.
These results also rationalize the use of a ``planar approximation"
in numerical simulations of FL depinning\cite{Enomoto}.

Another important consideration is the role of anisotropy in the bulk
material. It was recently shown that anisotropy leads to new
universality classes in interface depinning\cite{TKD}. We show
that this happens as well for FL depinning, in an even richer
fashion. The presence of a nonzero Hall angle affects the critical
behavior in a manner similar to anisotropy. These 
issues are discussed in more detail in Section~\ref{conclusion}.
 
The rest of the paper is organized as follows: In Section~\ref{eom},
we derive the general form of the equation of motion for a single
FL, starting from a reparametrization invariant (RI) descpription
of the FL dynamics. In Section~\ref{model},
we first establish the connection of Eqs.~(\ref{motion}) 
to the interface depinning problem for the simple case of an isotropic
medium with zero Hall angle. We then examine the linear response of the
system to derive the exponent identities (\ref{exp3}),(\ref{exp4}), and
(\ref{exp6}), which are later shown to be consistent with a formal 
RG treatment of the problem in more general circumstances.
In Section~\ref{formalism}, we present the MSR formalism and expand 
the generating functional around a 
self-consistent saddle point solution,  given by a mean-field 
theory. In Section~\ref{MFT}, we calculate response and 
connected correlation functions of the mean-field theory, which 
correspond to the bare propagators and vertex functions in 
a perturbative expansion. 
In Section~\ref{scaling}, we determine critical exponents through an 
$\epsilon$-expansion near $d=4$ dimensions, and in Sec.~\ref{numerical} 
we compare these with numerical results obtained by directly 
integrating the equations of motion. Finally, in Section~\ref{conclusion}
we discuss the physical significance of these results, the roles of nonlinear 
terms and anisotropy, and applicability of similar methods to related 
problems.

\section{Equations of Motion for a FL}
\label{eom}

In this section we derive a phenomenological 
equation that describes the coarse-grained (in space and time)
evolution of a single FL in a bulk type-II 
superconductor. The configuration of the FL at time $t$ is 
described by ${\vec R}(s,t)$, where $s$ is an arbitrary parameter
which we shall later equate to the $x-$component of ${\vec R}$.
The equations of motion are obtained by balancing the
``conservative" and ``dynamical" forces.
Conservative forces are derived from the energy functional
and depend only on the instantaneous configuration $\vec R(s)$ 
of the FL. They include the elastic force, random forces
due to the impurity potential $V$, and the Lorentz force
due to the bulk current. Dynamical forces, on the other 
hand, depend explicitly on the local velocity of the FL
and comprise the dissipative and Magnus forces\cite{ao}.

For notational simplicity, we set the external magnetic
field $\vec B$ along the $x$-axis and the
the average velocity $\vec v$ along ${\bf e_\parallel}$, suppressing 
the possible dependence of parameters on the relative orientation 
of $\vec B$ and ${\bf e_\parallel}$ due to anisotropy in the 
underlying material.
Such complications will be taken up later in Sec.~\ref{conclusion}.
An important consideration is the requirement that
the equation of motion be invariant under an 
arbitrary reparametrization ${\vec R}(s)\to{\vec R}(s')$ of the curve.
One such reparametrization invariant quantity is
the infinitesimal arclength $dl=ds\sqrt{g}$,
where $g\equiv\partial_s{\vec R}\cdot\partial_s{\vec R}$ is the metric.
The only physically observable motions 
of the FL are orthogonal to the local unit tangent vector
\[{\hat t}={1\over\sqrt{g}}\partial_s{\vec R}.\]
Assuming that the FL motion is overdamped, the conservative
force $\vec F_T$, which is derived below, is balanced 
by dynamical forces that are proportional to the local 
normal velocity 
$\vec v_n={\cal P}\cdot\partial_t\vec R
=\partial_t\vec R-(\partial_t\vec R\cdot\hat t)\hat t$.
(Here, ${\cal P}_{ij}\equiv\delta_{ij}-\hat t_i \hat t_j$ projects
any vector onto the local normal plane.)
Dynamical forces are not necessarily parallel to $\vec v_n$: 
In general, there is an angle $\phi$ (called the Hall angle) 
between the applied force and the velocity of the FL. Physically, 
this is due to the Magnus force which is orthogonal
to the velocity, and the Hall effect in the normal core 
of the FL\cite{Tinkham}. 
The equation of motion can then be written as
\begin{equation}
\label{fullmotion}
\eta{\cal P}\cdot\left\{\cos\phi\,\partial_t{\vec R}+
\sin\phi(\partial_t{\vec R})
\times{\hat t}\right\}={\vec F_T}.
\end{equation}

To determine the conservative force $\vec F_T$, 
consider the energy cost associated with a particular coarse-grained
configuration ${\vec R}(s)$ of the FL in the absence of a bulk
current, which is
\begin{equation}
E\left[{\vec R}(s)\right]=\int ds\sqrt{g}
\left\{\frac{\partial_s{\vec R}\cdot{\mbox{\boldmath$\sigma$}}
\cdot\partial_s{\vec R}}{g}
+V({\vec R}(s))\right\}.
\end{equation}
In the above equation, the symmetric tensor ${\mbox{\boldmath$\sigma$}}$ 
gives the core energy 
per unit length of the FL, and can be nondiagonal for an 
anisotropic sample. (Anharmonic contributions to the core energy 
can be ignored in a coarse-grained description and we will
systematically keep only the leading order elastic terms.)
The restoring force ${\vec F}_B$ is given by the 
energy cost of an infinitesimal virtual displacement
$\delta{\vec R}(s)$. After some rearrangement, we arrive at
\begin{eqnarray}
\delta E&=&-\int ds\sqrt{g}\, \delta \vec R \cdot{\cal P}\cdot\left\{
2{\mbox{\boldmath$\sigma$}}\cdot\vec\kappa
-\left(\hat t\cdot{\mbox{\boldmath$\sigma$}}\cdot\hat t\right)\vec\kappa
\phantom{\nabla_{{\vec R}} V({\vec R})}\right. \nonumber \\
& & \left. \qquad\qquad\qquad\qquad\qquad 
+V({\vec R})\vec\kappa -\nabla_{{\vec R}} V({\vec R})\right\} 
\nonumber \\
 &\equiv&-\int ds\sqrt{g}\, \delta{\vec R}\cdot{\vec F}_B,
\end{eqnarray}
where $\vec\kappa=g^{-1}{\cal P}\cdot\partial_s^2{\vec R}$ is 
the local curvature vector. To leading order, the random potential 
$V({\vec R})$ that multiplies $\vec\kappa$ can be approximated by its 
spatial average, and eliminated without loss of generality by choosing 
$\langle V\rangle=0$.
${\vec f}=-\nabla_{\vec R} V({\vec R})$ acts as a random force on each 
segment of the FL, whose correlations in general satisfy
\begin{equation}
\langle f_\alpha({\vec R})f_\gamma({\vec R}')\rangle=
\Delta_{\alpha\gamma}({\vec R}-{\vec R}').
\end{equation}
For now, we do not restrict the form of $\Delta$, apart from
the reasonable expectation that it decays quickly beyond a 
characteristic impurity size $a$.
When a bulk current ${\vec J}$ is present, the FL is also
subject to a Lorentz force ${\vec F}_L=\Phi_0{\vec J}\times{\hat t}$,
where $\Phi_0$ is the flux quantum. Thus, the total conservative force
acting on a section of the FL is given as 
\begin{equation}
{\vec F_T}={\cal P}\!\cdot\!
\left\{\frac{2{\mbox{\boldmath$\sigma$}}\cdot{\cal P}\cdot
\partial_s^2{\vec R}
-(\hat t\cdot{\mbox{\boldmath$\sigma$}}\cdot\hat t)\partial_s^2{\vec R}}{g}
+\Phi_0{\vec J}\times{\hat t}+\vec f\right\}.
\end{equation}
For an isotropic sample in the extreme type-II limit, 
the Bardeen-Stephen model gives\cite{Tinkham}
\begin{eqnarray*}
\sigma_{ij}&\approx&\delta_{ij}(\Phi_0/4\pi\lambda_s)^2
\ln(\xi_s/\lambda_s), \\ 
\eta&\approx&\Phi_0^2/(2\pi\xi_s^2c^2\rho_n), \\
\tan\phi&\approx&\rho_n/\rho_n^H, 
\end{eqnarray*}
where $\lambda_s$ is the London penetration depth, $\xi_s$ is the coherence
length, and $\rho_n, \rho_n^H$ are normal and Hall resistivities
of the non-superconducting core region, respectively. More general
expressions for these phenomenological parameters can be derived from a 
mesoscopic model based on a time-dependent 
Ginzburg-Landau theory\cite{Dorsey}.
 
Equation (\ref{fullmotion}) is highly nonlinear and generalizes 
those of ref.\cite{Tang} to the three-dimensional and anisotropic 
case. We now pick $\{\hat x,{\bf e_\parallel},{\bf e_\perp}\}$ as our 
coordinate axes,
and $x$ as the arbitrary parameter $s$, representing the FL as
${\vec R}(x,t)=x\hat x + {r_\parallel}(x,t){\bf e_\parallel}
+{r_\perp}(x,t){\bf e_\perp}$. In this
representation, $g=1+(\partial_x{r_\parallel})^2+(\partial_x{r_\perp})^2$,
$\vec J=J_\parallel{\bf e_\parallel}+J_\perp{\bf e_\perp}$, 
$\vec f=f_x\hat x+
{f_\parallel}{\bf e_\parallel}+{f_\perp}{\bf e_\perp}$, and 
\[
{\mbox{\boldmath$\sigma$}}=\pmatrix{\sigma_x&\sigma_{x\parallel}&
\sigma_{x\perp}\cr
\sigma_{x\parallel}&\sigma_\parallel&\sigma_\times\cr
\sigma_{x\perp}&\sigma_\times&\sigma_\perp}.
\] 
After some rearrangement, and elimination 
of higher-order terms coming from the elastic energy of the FL, 
we obtain the following evolution equations for the components 
${r_\parallel}$ and ${r_\perp}$: 

\end{multicols}
\widetext

\begin{mathletters}
\label{nleqn}
\begin{eqnarray}
{\eta\over\cos\phi}\partial_t{r_\parallel}&=&
[(2\sigma_\parallel-\sigma_x)-2\sigma_\times\tan\phi]
\partial_x^2{r_\parallel}
+[2\sigma_\times-(2\sigma_\perp-\sigma_x)\tan\phi]  
\partial_x^2{r_\perp} 
\nonumber \\
& &\
+ {\Phi_0\over\sqrt{g}}\left\{
J_\perp\left[1+(\partial_x{r_\parallel})^2\right]
-J_\parallel\left[\partial_x{r_\parallel}\partial_x{r_\perp}
-\tan\phi\sqrt{g}\right]
\right\} 
\nonumber \\
& &\
+{f_\parallel}\left[1+{\tan\phi\over\sqrt{g}}\partial_x{r_\parallel}
\partial_x{r_\perp}\right]
-{f_\perp}{\tan\phi\over\sqrt{g}}\left[1+(\partial_x{r_\parallel})^2\right]
-f_x\left[\partial_x{r_\parallel}-{\tan\phi\over\sqrt{g}}
\partial_x{r_\perp}\right], 
\\ \cr \nonumber \\
{\eta\over\cos\phi}\partial_t{r_\perp}&=&
[2\sigma_\times+(2\sigma_\parallel-\sigma_x)\tan\phi] 
\partial_x^2{r_\parallel}
+[(2\sigma_\perp-\sigma_x)+\sigma_\times\tan\phi]
\partial_x^2{r_\perp} 
\nonumber \\
& &\
+ {\Phi_0\over\sqrt{g}}\left\{
J_\perp\left[\partial_x{r_\parallel}\partial_x{r_\perp}
+\tan\phi\sqrt{g}\right]
-J_\parallel\left[1+(\partial_x{r_\perp})^2\right]
\right\} 
\nonumber \\
& &\ 
+{f_\parallel}{\tan\phi\over\sqrt{g}}\left[1+(\partial_x{r_\perp})^2\right]
+{f_\perp}\left[1-{\tan\phi\over\sqrt{g}}\partial_x{r_\parallel}
\partial_x{r_\perp}\right]
-f_x\left[\partial_x{r_\perp}+{\tan\phi\over\sqrt{g}}
\partial_x{r_\parallel}\right]. 
\end{eqnarray}
\end{mathletters}
These equations are clearly too complicated for an exhaustive analysis.
However, it is possible to perform a gradient expasion of the RHS of
Eqs.(\ref{nleqn}) when the fluctuations around the straight line are small,
i.e. $(\partial_x{r_\parallel})^2, (\partial_x{r_\perp})^2\ll 1$. 
In that case,
Eqs.(\ref{nleqn}) simplify to
\begin{mathletters}
\label{leqn}
\begin{eqnarray}
{\eta\over\cos\phi}\partial_t{r_\parallel}\!&=&\!
[(2\sigma_\parallel-\sigma_x)-2\sigma_\times\tan\phi]\partial_x^2{r_\parallel}
+[2\sigma_\times-(2\sigma_\perp-\sigma_x)\tan\phi]
\partial_x^2{r_\perp} \nonumber \\
& &\ + \Phi_0\!\left(J_\perp+J_\parallel\tan\phi\right)
\!+\!{f_\parallel}\!-\!{f_\perp}\tan\phi,\!\! \\ \cr \nonumber \\
{\eta\over\cos\phi}\partial_t{r_\perp}&=&
[2\sigma_\times+(2\sigma_\parallel-\sigma_x)\tan\phi]\partial_x^2{r_\parallel}
+[(2\sigma_\perp-\sigma_x)+\sigma_\times\tan\phi]
\partial_x^2{r_\perp}  \nonumber \\
& &\  + \Phi_0\!\left(
J_\perp\tan\phi-J_\parallel\right)\!+\!{f_\perp}\!-\!{f_\parallel}\tan\phi,
\end{eqnarray}
\end{mathletters}
neglecting all terms of $O\left((\partial_x{r_\parallel})^2,
(\partial_x{r_\perp})^2\right)$ or higher.

So far, we have not enforced the condition that ${\bf e_\parallel}$ 
points along the
average velocity of the FL. This is satisfied by the self-consistency
relation
\begin{equation}
\label{consistency}
\langle \partial_t{r_\perp}\rangle=0.
\end{equation} 
In the small fluctuation limit where Eqs.(\ref{leqn}) are valid, 
this condition is satisfied simply by setting 
$J_\parallel=J_\perp\tan\phi$. In order to study the scaling 
properties of this system in the framework of a field
theory, we generalize the FL to a manifold with $d$-dimensional
internal coordinates ${\bf x} \in \Re^d$. Further rearrangements, 
and addition of an infinitesimal external force 
${\mbox{\boldmath$\varepsilon$}}({\bf x},t)$ 
in order to study response functions, lead to
\begin{mathletters}
\label{eqB}
\begin{eqnarray}
\eta\partial_t{r_\parallel}=K_{11}\nabla_{\bf x}^2{r_\parallel}&+&
K_{12}\nabla_{\bf x}^2{r_\perp}
+F 
+
\tilde {f_\parallel}({\bf x},{\bf r}({\bf x},t))
+\varepsilon_1({\bf x},t), \\
\eta\partial_t{r_\perp}=K_{21}\nabla_{\bf x}^2{r_\parallel}&+&
K_{22}\nabla_{\bf x}^2{r_\perp}
+
\tilde {f_\perp}({\bf x},{\bf r}({\bf x},t))+\varepsilon_2({\bf x},t),
\end{eqnarray}
where $F=\Phi_0\sqrt{J_\parallel^2+J_\perp^2}=\Phi_0J$, and
\begin{eqnarray*}
\pmatrix{K_{11}& K_{12} \cr K_{21} & K_{22}} 
&=&
\pmatrix{\cos\phi & -\sin\phi \cr \sin\phi & \cos\phi} 
\pmatrix{2\sigma_\parallel-\sigma_x & 2\sigma_\times  \cr 
2\sigma_\times  & 2\sigma_\perp-\sigma_x }, \\ \\
\pmatrix{\tilde {f_\parallel} \cr \tilde {f_\perp}}  &=&
\pmatrix{\cos\phi & -\sin\phi \cr \sin\phi & \cos\phi} 
\pmatrix{{f_\parallel} \cr {f_\perp}}.
\end{eqnarray*}
The correlations of the random forces satisfy
\begin{equation}
\langle \tilde f_\alpha({\bf x},{\bf r})\tilde 
f_\gamma({\bf x},{\bf r}')\rangle=
\delta^d({\bf x}-{\bf x}')\tilde\Delta_{\alpha\gamma}({\bf r}-{\bf r}').
\end{equation}
\end{mathletters}
\begin{multicols}{2}
(Note that while both ${\bf r}$ and ${\bf x}$ are represented by bold
characters, ${\bf r}$ remains two dimensional, while ${\bf x}$ has
been promoted to a $d$-dimensional vector.)

In the special case of an isotropic medium with $\phi=0$, 
the equations further reduce to
\begin{mathletters}
\label{eqA}
\begin{eqnarray}
\eta\partial_t{r_\parallel}&=&
K\nabla_{\bf x}^2{r_\parallel}+F+{f_\parallel}({\bf x},
{\bf r}({\bf x},t))+\varepsilon_1({\bf x},t), \\
\eta\partial_t{r_\perp}&=&
K\nabla_{\bf x}^2{r_\perp}+{f_\perp}({\bf x},{\bf r}({\bf x},t))
+\varepsilon_2({\bf x},t),
\end{eqnarray}
where the correlations of the random forces satisfy
\begin{equation}
\langle f_\alpha({\bf x},{\bf r})f_\gamma({\bf x},{\bf r}')\rangle
=\delta_{\alpha\gamma}
\delta^d({\bf x}-{\bf x}')\Delta(|{\bf r}-{\bf r}'|).
\end{equation}
\end{mathletters}
We shall henceforth refer to Eqs.(\ref{eqA}) as Model A. 
Anisotropy and/or a nonzero Hall angle changes the scaling properties 
of the critical region, and we shall refer to this more general case, 
described by Eqs.(\ref{eqB}), as Model B.

\section{The Vector Depinning Model}
\label{model}

In this section, we study some properties of the system described
by Eqs.(\ref{eqB}) and (\ref{eqA}), in detail. Due to statistical 
translational symmetry in time $t$ and internal coordinates ${\bf x}$, 
we use the real $({\bf x},t)$ and Fourier $({\bf q},\omega)$ domains 
interchangeably when dealing with statistical averages. 

The vector depinning model differs from the CDW or interface
problems due to the presence of transverse fluctuations 
${r_\perp}({\bf x},t)$.
It is sometimes useful to recast the equations such that ${r_\perp}$ 
appears
as a function of ${r_\parallel}$ rather than $t$. The asymmetry in 
${r_\parallel}$ and
${r_\perp}$ occurs because ${r_\parallel}$ almost always moves in the 
forward 
direction\cite{backwards}, and therefore is a monotonous function of
$t$. Thus, for any particular realization of the random force
$f({\bf x},{\bf r})$, there is a unique point ${r_\perp}({\bf x},
{r_\parallel})$ that is visited
by the line for given coordinates $({\bf x},{r_\parallel})$. 
The evolution of 
${r_\perp}({\bf x},{r_\parallel})$ can be obtained schematically, 
by dividing
Eq.(\ref{eqB}b) by (\ref{eqB}a), as
\begin{equation}
\label{Astar}
\frac{\partial{r_\perp}}{\partial{r_\parallel}}=
\frac{K_{21}\nabla_{\bf x}^2{r_\parallel}+K_{22}\nabla_{\bf x}^2{r_\perp}
+\tilde{f_\perp}}
{K_{11}\nabla_{\bf x}^2{r_\parallel}+K_{12}\nabla_{\bf x}^2{r_\perp}
+\tilde{f_\parallel}+F}.
\end{equation}
We shall see that in most cases, the scaling properties of ${r_\perp}$
in relation to ${r_\parallel}$ can be obtained heuristically by inspecting 
Eq.(\ref{Astar}).

\subsection{Model A}

First of all, we establish the connection between Eq.(\ref{eqA})
and the interface depinning model for the special case of an isotropic 
system with $\phi=0$ (Model A). 
For a particular realization of randomness ${\bf f}({\bf x},{\bf r})$, 
Eq.(\ref{eqA}a) can be written as
\begin{equation}
\label{isomotion}
\eta\partial_t{r_\parallel}=K\nabla_{\bf x}^2{r_\parallel}
+f'\left({\bf x},{r_\parallel}({\bf x},t)\right)+F+\varepsilon_1({\bf x},t),
\end{equation}
where $f'({\bf x},{r_\parallel})={f_\parallel}\left({\bf x},
{r_\parallel},{r_\perp}({\bf x},{r_\parallel})\right)$ and 
${r_\perp}({\bf x},{r_\parallel})$ is determined by Eq.(\ref{Astar}).
It is quite plausible that, 
after averaging over all ${\bf f}$, the correlations in $f'$ will
also be short-ranged, albeit different from those of
${\bf f}$, since the dissipative dynamics will avoid maxima of the 
random potential, effectively reducing the average forces. 
In that case, the equation reduces exactly to the model studied by
NSTL and NF. Thus, the scaling of longitudinal fluctuations of the
FL near threshold will not change upon taking into account transverse
components, and the exponent relations (\ref{exp1}--\ref{exp4})
hold for Model A as well. We expect this argument to hold even
for Model B [Eqs.(\ref{eqB})] as long as 
$\nabla_{\bf x}^2{r_\perp}\ll\nabla_{\bf x}^2{r_\parallel}$, or when 
$\zeta_\perp<\zeta_\parallel$.

For the interface model, it is possible to show that $v(F)$ is a single 
valued function using the ``no passing rule" of Middleton\cite{Middleton}. 
The rule states that no interface (or CDW) can overtake another, if initially
every point on the first interface is behind the second one. This rule 
does not apply to the vector model: It is in principle possible
to have coexistence of moving and stationery FLs, allowing for the possibility 
of a discontinuous (multi-valued) ${\bf v}({\bf F})$. However, since a 
moving line 
samples an arbitrarily large region in the medium, it is plausible that the 
velocity self-averages at long times, resulting in a single 
valued ${\bf v}({\bf F})$ (i.e., no hysteresis). However, finite-size systems 
do suffer from such hysteresis which adversely affects numerical 
simulations of the model. These issues are further discussed in  
Sec.\ref{numerical}.

Several exponent identities can be deduced from the form of the 
linear response,
\begin{equation}
\chi_{\alpha\gamma}({\bf q},\omega)=\left<\frac{\partial r_\alpha({\bf q},
\omega)}
{\partial\varepsilon_\gamma({\bf q},\omega)}\right>,
\end{equation} 
in the $({\bf q},\omega)\to({\bf 0},0)$ limit. 
Due to the statistical symmetry of Eqs.(\ref{eqA}) under the 
transformation ${r_\perp} \to -{r_\perp}$, the linear response is diagonal.
Let us first set $\omega=0$ and examine the static response:  
An additional static force ${\mbox{\boldmath$\varepsilon$}}({\bf q})$ 
with zero 
spatial average (no ${\bf q}=0$ component) can be exactly compensated 
by the coordinate change 
\[
{\bf r}'_\alpha({\bf q},t)={\bf r}_\alpha({\bf q},t)
+(Kq^{2})^{-1}\varepsilon_\alpha({\bf q}).
\] 
The distribution of ${\bf f}$ does not change in the primed coordinates. 
Thus, the static linear response has the form
\begin{equation}
\label{staticresp}
\chi_{\alpha\gamma}({\bf q},\omega=0)=\delta_{\alpha\gamma}\frac{1}{Kq^{2}}.
\end{equation}
Since $\varepsilon_\parallel$ scales like the applied force,
the form of the linear response at the correlation length $\xi$
 gives the exponent identity 
\begin{equation}
\label{Aexp1}
\zeta_\parallel+1/\nu=2.
\end{equation}
Considering the transverse linear response 
seems to imply $\zeta_\perp=\zeta_\parallel$.  However, as will be
shown below, the static part of the transverse linear response becomes 
irrelevant at the critical RG fixed point, since 
$z_\perp>z_\parallel$. This is consistent
with the expectation that the dynamics is responsible for the
distinction between longitudinal and transverse modes.

\begin{figure}
\narrowtext
\epsfxsize=2.9truein
\centerline{\epsffile{\dir/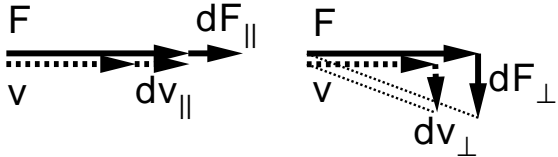}}
\centerline{ }
\caption{A graphical demonstration of Eqs.(\protect\ref{eqFl}-
\protect\ref{eqFt}).
When a longitudinal force is applied, the direction is not changed
and all changes are in the magnitude $F(v)$. For a transverse force, 
$F$ does not change to linear order in $dF_\perp$, but ${\bf v}$
changes direction to remain parallel to ${\bf F}$.} 
\label{dfdv}
\end{figure}

Why are the relaxational dynamics different in the two fluctuation
directions near depinning? The answer can be traced to a simple symmetry 
argument, which requires ${\bf F}$ and ${\bf v}$ to remain parallel, i.e.
\begin{equation}
\label{Fv}
{\bf F}({\bf v})=F(v){\hat v},
\end{equation}
where ${\hat v}={\bf v}/v$, and $F$ is some (scalar) function which depends on
only the magnitude $v$, of velocity. 
For small deviations around ${\bf v}=v{\bf e_\parallel}$, we thus obtain 
(see Fig.~\ref{dfdv})
\begin{eqnarray}
\label{eqFl}
\frac{\partial F_\parallel}{\partial v_\parallel} &=& \frac{dF}{dv}, \\
\label{eqFt}
\frac{\partial F_\perp}{\partial v_\perp} &=& \frac{F}{v}.
\end{eqnarray}

These two derivatives clearly scale differently in the $v\to 0$ limit,
which causes a separation of relaxation time scales, as shown below.

Now consider the response to a spatially
uniform (${\bf q}={\bf 0}$), but time-dependent, external force 
${\mbox{\boldmath$\varepsilon$}}(t)$. 
The leading term in the dynamic response is intricately connected to 
${\bf v}({\bf F})$: When a slowly varying uniform external force 
${\mbox{\boldmath$\varepsilon$}}(t)$ is 
applied, the FL responds as if the instantaneous external force 
${\bf F}+{\mbox{\boldmath$\varepsilon$}}$ is a constant, i.e. it moves 
with the average velocity
\begin{equation}
\langle\partial_t r_\alpha\rangle=v_\alpha({\bf F}+
{\mbox{\boldmath$\varepsilon$}})\approx 
v_\alpha({\bf F})+\frac{\partial v_\alpha}{\partial F_\gamma}
\varepsilon_\gamma.
\end{equation}
Therefore, near the depinning transition,
\begin{eqnarray}
\chi_\parallel({\bf q}=0,\omega) &\simeq& 
\frac{1}{-i\omega(dF/dv)+O(\omega^2)}, \\
\chi_\perp({\bf q}=0,\omega) &\simeq& \frac{1}{-i\omega(F_c/v)+O(\omega^2)}.
\end{eqnarray}

Eq.(\ref{staticresp}) can be combined with the above to yield
a Taylor expansion of the inverse linear response around 
$({\bf q},\omega)=({\bf 0},0)$ that reads
\begin{eqnarray}
\chi_\parallel^{-1}({\bf q},\omega) &\simeq& Kq^2-i\omega(dF/dv)
+{\rm h.o.t.}, \\
\chi_\perp^{-1}({\bf q},\omega) &\simeq&Kq^2-i\omega(F_c/v)+{\rm h.o.t.}
\end{eqnarray}
The zero of $\chi^{-1}$ in the complex $\omega-$plane for a given value of
the wavevector ${\bf q}$ gives the relaxation time of the corresponding
mode. Thus, the relaxation times of fluctuations with wavelength $\xi$ are
\begin{eqnarray}
\tau_\parallel(q=\xi^{-1}) &\sim& \left(q^2\frac{dv}{dF}\right)^{-1}
\sim\xi^{2+(\beta-1)/\nu}\sim\xi^{z_\parallel}, \\
\tau_\perp(q=\xi^{-1}) &\sim& \left(q^2\frac{v}{F_c}\right)^{-1}
\sim\xi^{2+\beta/\nu}\sim\xi^{z_\perp},
\end{eqnarray}
which in turn yield
\begin{eqnarray}
\label{Aexp2}
\beta&=&1+(z_\parallel-2)\nu, \\
\label{Aexp3}
z_\perp&=&z_\parallel+1/\nu.
\end{eqnarray}
Thus,  $z_\perp>z_\parallel$ as noted earlier. This difference arises
entirely from the different scaling properties of 
$dv/dF$ $[\sim(F-F_c)^{\beta-1}]$ and  $v/F$ $[\sim(F-F_c)^\beta]$
near the depinning transition, as noted earlier.

\subsection{Model B}

A similar linear response analysis can be made for the more general
case of Model B. The leading contributions to the static and
dynamic part of the inverse linear response are given by
\begin{eqnarray}
\chi^{-1}_{\alpha\gamma}({\bf q},\omega=0)&=& K_{\alpha\gamma}q^{2}, \\
\chi^{-1}_{\alpha\gamma}({\bf q}={\bf 0},\omega)&=& -i\omega
(\partial v_\alpha / \partial F_\gamma)^{-1}. 
\end{eqnarray}
The relation between the external force and the drift velocity can
in general be written as
\begin{equation}
\label{Fofv}
{\bf F}({\bf v})=F(v,\theta)\left[\cos\phi(v,\theta){\hat v}+
\sin\phi(v,\theta){\hat x}\times{\hat v}\right].
\end{equation}
Both $F$ and $\phi$ in general depend on the orientation of ${\hat v}$, 
parametrized by an angle $\theta$ in the $yz$-plane. 
Then, for small deviations around ${\bf v}=v{\bf e_\parallel}$, 
\begin{equation}
\label{dFdv}
\pmatrix{dF_\parallel \cr \cr dF_\perp \cr} =\pmatrix{
A_{11}
& -\frac{1}{v}A_{12} \cr \cr
A_{21} 
& \frac{1}{v}A_{22} \cr}
\pmatrix{dv_\parallel \cr \cr dv_\perp \cr},
\end{equation}
where
\begin{eqnarray*}
A_{11}&=&\partial_v (F\cos\phi), \\
A_{21}&=&\partial_v(F\sin\phi), \\
A_{12}&=&F\sin\phi-\partial_\theta(F\cos\phi), \\
A_{22}&=&F\cos\phi+ \partial_\theta(F\sin\phi).
\end{eqnarray*}
The scaling of diagonal elements in the linear response are the
same as in Model A. Therefore, exponent identities 
(\ref{Aexp2}-\ref{Aexp3}) hold in the more general case of Model B as well. 

\section{MSR Formalism}
\label{formalism}

We use the formalism of MSR\cite{MSR} to compute response 
and correlation functions for the dynamical system described by 
Eqs. (\ref{eqB}). After some rearranging, we obtain
\begin{eqnarray}
\label{discmotion}
\lefteqn{\eta\frac{\partial r_\alpha({\bf x},t)}{\partial t} 
= \int{d^d\!x\,}'\,J_{\alpha\gamma}({\bf x}-{\bf x}')
r_\gamma({\bf x}',t)}\quad & & 
\nonumber \\
& &-r_\alpha({\bf x},t)+\tilde f_\alpha\left({\bf x},
{\bf r}({\bf x},t)\right)+F_\alpha
+\varepsilon_\alpha({\bf x},t),
\end{eqnarray}
where the tensor {\bf J} is given by its Fourier transform as
$J_{\alpha\gamma}({\bf q})=\delta_{\alpha\gamma}-K_{\alpha\gamma}q^2$.
Introducing an auxiliary field ${\bf \hat r}({\bf x},t)$, the generating 
functional $Z$ is given by
\begin{equation}
\label{Z}
Z=\int{[d{\bf r}][d{\bf \hat r}]{\cal J}[{\bf r}]\exp(S)},
\end{equation}
where
\begin{eqnarray}
\lefteqn{S=i\int{d^d\!x\,}{dt\,}{\hat r}_\alpha({\bf x},t)\left\{\eta
\partial_t r_\gamma -K_{\alpha\gamma}\nabla_{\bf x}^2 
r_\gamma \right.}\quad & &\nonumber \\
& &\left.-F_\alpha-\tilde f_\alpha\left({\bf x},{\bf r}({\bf x},t)\right)
-\varepsilon_\alpha({\bf x},t)\right\}.
\end{eqnarray}

Clearly, this coarse-grained continuum picture of the system breaks down
at length scales shorter than the core radius of the FL. 
Therefore, there is a natural cutoff $\Lambda$ in {\bf q}-space for the 
functional integrals in Eq.~(\ref{Z}). $Z$ can be used to generate response 
and  correlation functions of ${\bf r}$, since integrating over 
${\bf \hat r}$ gives 
delta functions that impose the solution to the equation of motion 
(\ref{discmotion}). The Jacobian ${\cal J}[{\bf r}]$ fixes the renormalization 
of $Z$ such that the delta functions integrate to unity, and will be 
suppressed henceforth. Since $Z=1$ independent of the realization of 
randomness, response and correlation functions can also be generated 
using the disorder-averaged generating function $\overline Z=\int
{[d{\bf r}][d{\bf \hat r}]\langle\exp(S)\rangle}$. For example, the two-point 
correlation function is given by
\[
\langle r_\alpha({\bf x},t)r_\gamma({\bf x}',t')\rangle 
= \int{[d{\bf r}][d{\bf \hat r}]r_\alpha({\bf x},t)r_\gamma({\bf x}',t')
\langle\exp(S)\rangle},
\]
and the linear response is
\[
\left<\frac{\delta r_\alpha({\bf x},t)}{\delta\varepsilon_\gamma({\bf x}',t')}
\right>= -i\int{[d{\bf r}][d{\bf \hat r}]r_\alpha({\bf x},t)
{\hat r}_\gamma({\bf x}',t')
\langle\exp(S)\rangle}.
\]
In order proceed, we discretize in ${\bf x}$-space: ${\bf r}({\bf x},t)
\to{\bf r}_i(t)$.
Introducing two conjugate fields ${\bf R}_i(t),{\bf \hat R}_i(t)$, 
$\overline Z$ 
can be rewritten as 
\begin{eqnarray}
{\overline Z} &=& \int[d{\bf R}][d{\bf \hat R}]\exp(\tilde S), \\
\label{S}
\tilde S &=& \sum_j \ln {\overline Z_j}\left({\bf R}_j,
{\bf \hat R}_j\right) \nonumber \\
& &\quad-i\int{dt\,}\sum_{i,j}
{\bf \hat R}_i(t)\cdot{\bf J}^{-1}_{ij}\cdot{\bf R}_j(t),
\end{eqnarray}
where ${\overline Z_j}\left({\bf R}_j,{\bf \hat R}_j\right)$ is given by 
\begin{eqnarray}
{\overline Z_j}&=&\int{[d{\bf r}_j][d{\bf \hat r}_j]
\left<\exp\int{dt\,}\left[i{\bf \hat R}_j(t)
\cdot{\bf r}_j(t) \right.\right.} \\
& &\quad +i{\bf \hat r}_j(t)\cdot
\left\{\eta\partial_t{\bf r}_j(t)-{\bf R}_j(t) \right. \nonumber \\
& &\left.{\phantom\int}\left.{\phantom{\bf \hat R}}\left.
+{\bf r}_j(t)-\tilde{\bf f}_j\left({\bf r}_j(t)\right)-{\bf F}
-{\mbox{\boldmath$\varepsilon$}}_j(t)\right\}\right]\right>. \nonumber
\end{eqnarray}
Note that this factorization of the disorder-dependent part of the action
to local functionals ${\overline Z_j}$ is possible only if the random
forces $\tilde{\bf f}_j$ are independent at each site $j$, as assumed in 
Eq.~(\ref{statistics}). 
${\overline Z_j}$ can be evaluated by an expansion around the saddle-point 
approximation. The integrand of the exponential is a maximum when, 
for each $j$,
\begin{eqnarray*}
-\sum_i{\bf J}_{ij}^{-1}\cdot{\bf \hat R}_i^0-
\left<{\bf \hat r}_j\right>_{\rm MF}={\bf 0}, \\
-\sum_i{\bf J}_{ij}^{-1}\cdot{\bf R}_i^0+\left<{\bf r}_j\right>_{\rm MF}
={\bf 0}, 
\end{eqnarray*}
which has a solution ${\bf \hat R}^0_j={\bf 0},\;{\bf R}^0_j={\bf v} t$ 
for all $j$. 
Here, {\bf v} is determined self-consistently as a function of {\bf F}\ by 
requiring $\left<{\bf r}_j\right>_{\rm MF}={\bf v} t$, where averages 
$(\left<\dots\right>_{\rm MF})$
are generated from ${\overline Z_j}$ evaluated at the saddle point, which
is identical for each $j$:
\begin{eqnarray}
\label{ZMF}
\lefteqn{Z_{\rm MF}=\int[d{\bf r}_j][d{\bf \hat r}_j]\left<\exp i\int dt\;
{\bf \hat r}_j(t) 
\right.} \quad & &  \\
& & \left.{\phantom\int}\cdot\left\{{\eta\partial_t 
{\bf r}_j}(t)-{\bf v} t +{\bf r}_j(t)-
\tilde{\bf f}_j\left({\bf r}_j(t)\right)
-{\bf F}
-{\mbox{\boldmath$\varepsilon$}}_j(t)\right\}\right>. \nonumber 
\end{eqnarray}
$Z_{\rm MF}$ can be identified as the MRS generating function 
for a mean-field (MF) approximation to Eq.~(\ref{discmotion}), 
obtained by setting $J_{\alpha\gamma}({\bf x}-{\bf x}')
=\delta_{\alpha\gamma}N^{-1}$, 
where $N=\int d^dx$. (The first
term in the RHS of (\ref{discmotion}) is then self-consistently
equal to $\langle{\bf r}\rangle_{\rm MF}={\bf v} t$.)
Redefining the field variables as ${\bf R} \to {\bf R}+{\bf v} t,\,
i{\bf \hat R}\to{\bf \hat R}$ (for notational simplicity), the expansion for 
$\ln {\overline Z_j}\left({\bf R}_j,{\bf \hat R}_j\right)$ is given by 
\begin{eqnarray}
\label{lnzj}
\lefteqn{\ln {\overline Z_j}\left({\bf R}_j,{\bf \hat R}_j\right) = 
\sum_{\{m_\alpha,n_\alpha\}=0}^{\infty}
\left(\prod_{\alpha}\frac{1}{m_\alpha!n_\alpha!}\right)} 
\quad & & \nonumber \\
& &\times \int  \prod_\alpha \left\{
\prod_{s_\alpha=1}^{m_\alpha}dt_{\alpha s_\alpha}
{\hat R}_{j\alpha}(t_{\alpha s_\alpha}) 
\prod_{s_\alpha'=1}^{n_\alpha}dt_{\alpha s_\alpha'}' 
R_{j\alpha}(t_{\alpha s_\alpha'}')\right\} \nonumber \\
& &\qquad\qquad\qquad\qquad\times{\cal V}_{\{m_\alpha,n_\alpha\}}
\left(\{t_{\alpha s_\alpha}\};\{t_{\alpha s_\alpha'}'\}\right).
\end{eqnarray}
The vertex functions ${\cal V}$ are obtained by evaluating
derivatives of\ $\ln {\overline Z_j}$ with respect to the fields 
{\it at the saddle point}, and are given precisely by
connected correlation and response functions of the MF system
decribed by Eq.~(\ref{ZMF}):
\begin{eqnarray}
\label{vertex}
\lefteqn{{\cal V}_{\{m_\alpha,n_\alpha\}}\left(\{t_{\alpha s_\alpha}\};
\{t_{\alpha s_\alpha'}'\}\right)} \quad & & \\
&=& \left[\prod_\alpha\prod_{s_\alpha'=1}^{n_\alpha} 
\frac{\partial}{\partial \varepsilon_{j\alpha}(t_{\alpha s_\alpha'}')}
\right]
\left<\prod_\alpha\prod_{s_\alpha=1}^{m_\alpha}
r_{j\alpha}(t_{\alpha s_\alpha})\right>_{{\rm MF},c}. \nonumber
\end{eqnarray}
Thus, once the mean-field system is solved, 
correlation functions of ${\bf R},{\bf \hat R}$ can be studied
through a momentum space RG treatment to obtain the 
scaling exponents of the fields in the long-time,
large wavelength (hydrodynamic) limit. ${\bf R}$ and ${\bf \hat R}$ 
are like
coarse-grained forms of the original fields ${\bf r}$ and ${\bf \hat r}$ 
since
all correlation functions of ${\bf r},{\bf \hat r}$ are  
equal to corresponding correlation functions of ${\bf R},{\bf \hat R}$
 in the hydrodynamic limit\cite{NF92}. Therefore, it is sufficient
to find the scaling behavior of ${\bf R},{\bf \hat R}$ to deduce the desired
critical exponents.

\section{Mean Field Theory}
\label{MFT}

In this section, we calculate response and correlation
functions of the local system described by $Z_{\rm MF}$, 
which gives the vertex functions in the diagrammatic
expansion of $\tilde S$. We will only need to calculate 
the leading terms as higher order vertices will turn out to be
irrelevant in the critical region. Due to the averaging,
$Z_{\rm MF}$ is identical at all sites $j$, and it is sufficient to
examine a single point. Setting $\overline{\bf r}(t)
\equiv{\bf r}_j(t)-{\bf v} t$, 
and ${\mbox{\boldmath$\varepsilon$}}(t)
\equiv{\mbox{\boldmath$\varepsilon$}}_j(t)$, the equation of motion 
becomes

\begin{figure}
\narrowtext
\epsfxsize=2.9truein
\centerline{\epsffile{\dir/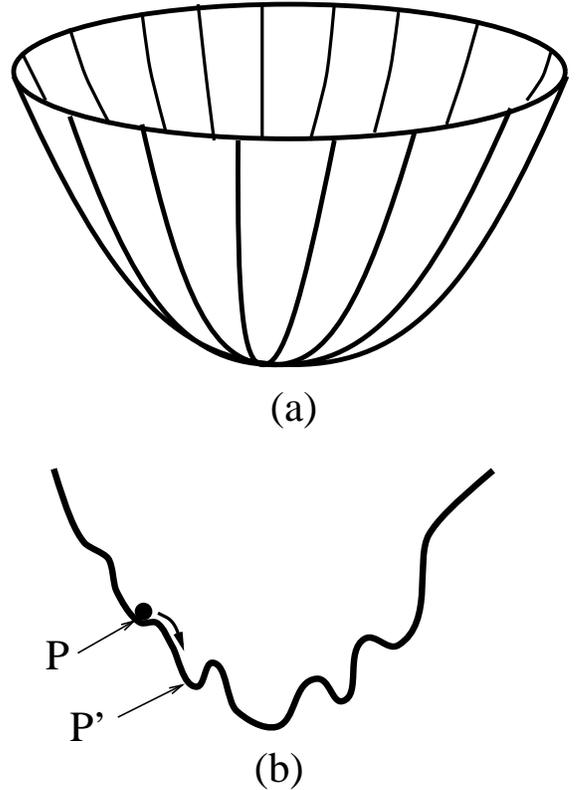}}
\caption{(a) The effective potential $V_{\rm eff}$. The 
random part (not shown) superimposed on the paraboloid slides with 
velocity $-{\bf v}$. (b) A cross section of $V_{\rm eff}$. The particle 
stays in 
a local minimum $P$ for a time of $O(v^{-1})$, after
which the minimum disappears and the particle finds another local
minimum $P'$ within a finite time. Time averages are dominated
by the slow portion of the motion as $v\to 0$.}
\label{Veff}
\end{figure}

\begin{equation}
\label{MFmotion}
\eta\left(\frac{d\overline r_\alpha}{dt}+v_\alpha\right) 
= -\overline r_\alpha +F_\alpha+
\tilde f_\alpha\left({\bf v} t+\overline{\bf r}(t)\right)
+\varepsilon_\alpha(t).
\end{equation}
${\bf F}$ is determined as a function of ${\bf v}$ self-consistently 
by requiring
that $\langle\overline{\bf r}\rangle_{\rm MF}={\bf 0}$.
The scaling behavior of ${\bf F}_{\rm MF}({\bf v})$ near threshold 
can be determined
from the following argument:
For $v\ll\eta^{-1}F$, the particle follows a local minimum $P$
of the effective potential 
\[
V_{\bf eff}(\overline{\bf r},t)=
V\left({\bf x},{\bf v} t+\overline{\bf r}(t)\right)
+\frac{|\overline{\bf r}(t)|^2}{2}
-{\bf F}\cdot\overline{\bf r}(t).
\]
A representative snapshot of $V_{\bf eff}$, which consists
of a paraboloid centered at $\overline{\bf r}={\bf F}$ with a superimposed 
random potential,
is shown in Fig.~\ref{Veff}. The position of the local minimum $P$ 
shifts with a velocity of $O(v)$ as time progresses. Eventually, 
$P$ disappears at a saddle point as it is pushed up the sides of 
the hyperparaboloid. At this moment, the particle quickly moves to a 
new local minimum $P'$, after which it starts following the slow 
motion of $P'$, as shown in Fig.~\ref{Veff}. 
For scalloped random potentials with discontinuous derivatives 
at the saddle points, the particle starts moving with a velocity of 
$O(1)$ (i.e., independent of $v$ as $v\to0$) {\it as soon as} $P$ 
disappears, and reaches the vicinity of
$P'$ in $O(1)$ time, giving the result $\beta_{\rm MF}=1$. (In contrast,
for smooth potentials, there is a $v$-dependent acceleration time 
just after $P$ disappears, which contributes to the critical dynamics and 
gives $\beta_{\rm MF}=3/2.$\cite{Fisher85,Middleton}.) We have also 
numerically integrated Eq.~(\ref{MFmotion}) (for Model A) 
to verify that $\beta_{\rm MF}=1$. 

Next, we proceed to compute vertex functions 
${\cal V}_{\{m_\alpha,n_\alpha\}}
\left(\{t_{\alpha s_\alpha}\};\{t_{\alpha s_\alpha'}'\}\right)$ 
in the perturbative expansion of $\tilde S$, which correspond to
response and connected correlation functions of the MF theory,
in increasing order in the field variables ${\bf R},{\bf \hat R}$. From now
on, we set 
$m=m_\parallel+m_\perp$, and $n=n_\parallel+n_\perp$.

\subsection{Average position (m=1, n=0)}
 
By construction $\left<\overline{\bf r}\right>_{\rm MF}=0$, but we prefer 
to expand
around the true ${\bf F}({\bf v})$ instead of the mean-field value of the
force ${\bf F}_{\rm MF}({\bf v})$. Since the effect of an additional uniform 
static force ${\bf F}-{\bf F}_{\rm MF}({\bf v})$ can be fully counteracted 
by a shift in $\overline{\bf r}$, this does not affect connected 
correlation or response functions. Thus, the only effect of this 
shift is to produce an additional term 
$\sum_i[{\bf F}-{\bf F}_{\rm MF}({\bf v})]\cdot{\bf \hat R}_i$ 
in $\tilde S$, which only
has a $q=0$ component and does not  directly enter the 
renormalization of higher order terms. 

\subsection{Linear Response (m=1, n=1)}

The linear response is given by the rank 2 tensor,
\[
{\tilde\chi}_{\alpha\gamma}(t-t')=
\left<\frac{\delta \overline r_\alpha(t)}{
\delta\varepsilon_\gamma(t')}\right>_{\rm MF}.
\]
We are only interested in the low-frequency form of the Fourier
transformed linear response ${\tilde{\mbox{\boldmath$\chi$}}}(\omega)$, 
i.e. when ${\mbox{\boldmath$\varepsilon$}}$ 
is slowly varying in time. In this case, we can write
${\mbox{\boldmath$\varepsilon$}}(t)\approx{\mbox{\boldmath$\varepsilon$}}_0
+{\mbox{\boldmath$\varepsilon$}}_1t$, neglecting terms proportional to 
$\ddot{\mbox{\boldmath$\varepsilon$}}$. To find the response 
$\overline{\bf r}_{\mbox{\boldmath$\varepsilon$}}(t)$, let us define
\[
\overline{\bf r}_{\mbox{\boldmath$\varepsilon$}}'(t)
=\overline{\bf r}_{\mbox{\boldmath$\varepsilon$}}(t)
-{\mbox{\boldmath$\varepsilon$}}_0-{\mbox{\boldmath$\varepsilon$}}_1 
t-{\bf F}_{\rm MF}({\bf v})+{\bf F}_{\rm MF}({\bf v}
+{\mbox{\boldmath$\varepsilon$}}_1).
\]
Taking a time derivative and using Eq.(\ref{MFmotion}),
we obtain
\begin{eqnarray}
\eta(\dot{\overline{\bf r}}'_{\mbox{\boldmath$\varepsilon$}}+{\bf v}
+{\mbox{\boldmath$\varepsilon$}}_1)
= -\overline{\bf r}'_{\mbox{\boldmath$\varepsilon$}}
&+&{\bf F}_{\rm MF}({\bf v}+{\mbox{\boldmath$\varepsilon$}}_1) \nonumber \\
&+&\tilde {\bf f}(({\bf v}+{\mbox{\boldmath$\varepsilon$}}_1)t
+\overline{\bf r}_{\mbox{\boldmath$\varepsilon$}}'
-{\bf F}_{\mbox{\boldmath$\varepsilon$}}),
\end{eqnarray}
where ${\bf F}_{\mbox{\boldmath$\varepsilon$}}={\bf F}_{\rm MF}({\bf v}
+{\mbox{\boldmath$\varepsilon$}}_1)-{\bf F}_{\rm MF}({\bf v})
-{\mbox{\boldmath$\varepsilon$}}_0$.
But now, $\left<\overline{\bf r}'_{\mbox{\boldmath$\varepsilon$}}\right>
={\bf 0}$ 
by definition of 
${\bf F}_{\rm MF}$. (The random force $\tilde{\bf f}$ is evaluated at 
points shifted by a constant amount ${\bf F}_{\mbox{\boldmath$\varepsilon$}}$,
but this has no significance upon averaging over randomness.) 
This gives
\begin{equation}
\label{response}
\langle\overline{\bf r}_{\mbox{\boldmath$\varepsilon$}}(t)\rangle
={\mbox{\boldmath$\varepsilon$}}(t)+{\bf F}_{\rm MF}({\bf v})-
{\bf F}_{\rm MF}\left({\bf v}+\dot{\mbox{\boldmath$\varepsilon$}}(t)\right)
+O(\ddot{\mbox{\boldmath$\varepsilon$}}).
\end{equation} 
Expanding ${\bf F}_{\rm MF}({\bf v}+\dot{\mbox{\boldmath$\varepsilon$}})$ 
for small $\dot{\mbox{\boldmath$\varepsilon$}}$,
we obtain
\begin{equation}
{\tilde\chi}_{\alpha\gamma}(\omega)=\delta_{\alpha\gamma}
+i\omega\left[\frac{\partial F_{{\rm MF}\alpha}({\bf v})}{\partial v_\gamma}
\right]_{{\bf v}=v{\bf e_\parallel}}+O(\omega^2).
\end{equation}
Since $\beta_{\rm MF}=1$, the linear response tensor will have the form
\begin{equation}
\label{linresp}
\tilde {\mbox{\boldmath$\chi$}}(\omega)={\bf 1}+i\omega\pmatrix{
A_{11} & -\frac{1}{v}A_{12} \cr \cr
A_{21} & \frac{1}{v}A_{22} \cr},
\end{equation}
where $A_{\alpha\gamma}$ approach constants as $v\to 0$
(cf. Eq.(\ref{dFdv})). For Model A, $\tilde{\mbox{\boldmath$\chi$}}(\omega)$ 
is diagonal
due to symmetry, and $A_{12}=A_{21}=0$. 

\subsection{Nonlinear response (m=1, n$>$1)}

Assuming that ${\bf F}_{\rm MF}({\bf v})$ has a Taylor expansion around 
${\bf v}=v{\bf e_\parallel}$
for $v>0$, we can expand the RHS of Eq.~(\ref{response}) to obtain the 
nonlinear response of the model. The leading term in the low-frequency 
limit is proportional to $\omega^n$, and it is straightforward to show 
that the contribution of these terms 
to $\tilde S$ is 
\begin{eqnarray}
\label{m1series}
-\frac{1}{n_\parallel!n_\perp!}\int{d^d\!x\,}&{dt\,}&
\left[\frac{\partial^n F_\alpha({\bf v})}{\partial^{n_\parallel}v_\parallel
\partial^{n_\perp}v_\perp}\right]_{{\bf v}=v{\bf e_\parallel}}
 \nonumber \\
& & \times{\hat R}_\alpha({\bf x},t)(\partial_tR_\parallel)^{n_\parallel}
(\partial_tR_{\perp})^{n_{\perp}}.
\end{eqnarray}
These terms are irrelevant at the RG fixed point, as we shall show later.

\subsection{Two-point Correlation Functions (m=2, n$\geq$0)}

At low velocities, the particle spends most of the time near a local
minimum, jumping abruptly to the next one when this minimum
disappears. Therefore, the time scale associated with the correlation
functions is given by the temporal separation between two consecutive jumps,
which scales as $1/v$. In the $v\to 0$ limit, the correlation 
functions depend on $t$ only through the rescaled time variable 
$u\equiv vt$, since the positions of successive minima near threshold
are determined by energetic considerations, and do not depend on $v$. 
(The correlation functions may also depend on the drift direction ${\hat v}$. 
We shall suppress this dependence for notational brevity.) 
Let us define 
\begin{equation}
\left<\overline r_\alpha(t)\overline r_\gamma(t')\right>_{{\rm MF},c}
\equiv C_{\alpha\gamma}\left(v(t-t')\right).
\end{equation}
Since successive positions of the local minima are uncorrelated,
 we expect that $C_{\alpha\gamma}(u)$ decay quickly 
as a function of $u\equiv vt$ for $|u|>1$. 
By definition, 
\begin{eqnarray*}
C_\parallel(u)&\equiv&C_{11}(u)=C_{11}(-u), \\
C_\perp(u)&\equiv&C_{22}(u)=C_{22}(-u), \\
{1\over 2}C_\times(u)&\equiv&C_{12}(u)=C_{21}(-u).
\end{eqnarray*}
As a result of the abrupt jumps from one minimum to another, 
$C_{\alpha\gamma}(vt)$ have a discontinuous derivative at the origin, 
rounded at a scale of $O(v)$. In Model A, $C_\times(u)=0$ due to symmetry.

The only other important terms in the effective action $\tilde S$ 
involve the series $m=2, n=n_\parallel>0, n_\perp=0$. All vertex functions
associated with this series are given by the response of connected
correlation functions to {\it longitudinal} forces. These response
functions are intimately related to the two-point correlation functions
$C_{\alpha\gamma}(u)$ by the following argument: Static forces only change 
linear response, and do not affect connected correlation functions. 
For a slowly varying external force
$\varepsilon(t){\bf e_\parallel}$, however, the system will respond as if the 
instantaneus velocity is $(v+\dot\varepsilon){\bf e_\parallel}$. Neglecting
terms proportional to $\ddot\varepsilon$,
\begin{eqnarray*}
\left[\left<\overline r_\alpha(t)\overline r_\gamma(t')\right>_{{\rm MF},c}
\right]_{\varepsilon}
&=&C_{\alpha\gamma}\left((v+\dot\varepsilon)(t-t')\right)
+O(\ddot\varepsilon) \nonumber \\
&\approx&C_{\alpha\gamma}\left(v(t-t')+\varepsilon(t)-\varepsilon(t')\right).
\end{eqnarray*}
Now, Taylor expanding $C_{\alpha\gamma}$ around $v(t-t')$ 
and taking successive functional derivatives with respect to 
$\varepsilon$, we finally obtain the contribution of this series
to $\tilde S$ as
\begin{eqnarray}
\label{m2series}
{\cal U}=\lefteqn{\sum_{n=1}^\infty\frac{1}{2!n!} 
\int{d^d\!x\,}{dt\,} dt'\,{\hat R}_\alpha({\bf x},t)
{\hat R}_\gamma({\bf x},t')} 
\quad & & \\
& &\qquad\times U_{\alpha\gamma,n}\left(v(t-t')\right)
\left[R_\parallel({\bf x},t)-R_\parallel({\bf x},t')\right]^n,  \nonumber 
\end{eqnarray}
where $U_{\alpha\gamma,n}(u)$ is the $n$th derivative of 
$C_{\alpha\gamma}(u)$. 

The vertices with  $m=2,n_\perp>0$ and $m>3$ are all irrelevant, 
as shown in the next section. 

\end{multicols}
\widetext

\section{Scaling and RG}
\label{scaling}

The terms in $\tilde S$ that are up to second order in the fields are
\begin{eqnarray}
\label{quadratic}
{\tilde S}_0&=&-\int{dt\,}{d^d\!x\,} [{\bf F}-{\bf F}_{\rm MF}({\bf v})]
\cdot{\bf \hat R}({\bf x},t) 
\nonumber \\
& & \ -\frac{1}{2}\int\limits_{{\bf q},\omega} 
\left[\matrix{{\bf \hat R}(-{\bf q},-\omega) \cr \cr 
 {\bf R}(-{\bf q},-\omega)\cr }\right]^{\rm T} 
\!\!\!\cdot\!\left[\matrix{-{\bf C}(\omega) & 
{\bf J}^{-1}({\bf q})-\tilde{\mbox{\boldmath$\chi$}}(\omega) \cr \cr
{\bf J}^{-1}(-{\bf q})-\tilde{\mbox{\boldmath$\chi$}}(-\omega) 
& {\bf 0} \cr}\right]\!\cdot\!
\left[\matrix{{\bf \hat R}({\bf q},\omega) \cr \cr 
{\bf R}({\bf q},\omega) }\right] , 
\end{eqnarray}
where $J^{-1}_{\alpha\gamma}({\bf q})=
\left(\delta_{\alpha\gamma}-K_{\alpha\gamma}q^2\right)^{-1}
\approx \delta_{\alpha\gamma}+K_{\alpha\gamma}q^2$ for small $q$. 
For notational brevity, we use $\int_{{\bf q},\omega}$ to denote
$\int{{d^d q \over (2\pi)^d}\,}{{d\omega \over 2\pi}\,}$. 
Using Eq.(\ref{linresp}), the quadratic form in 
the action can be written as 
\begin{equation}
\label{qform}
-{1\over 2}\int\limits_{{\bf q},\omega}\left[\matrix{
{\hat R}_\parallel(-{\bf q},-\omega) \cr  
{\hat R}_\perp(-{\bf q},-\omega) \cr 
R_\parallel(-{\bf q},-\omega) \cr 
{1\over v}R_\perp(-{\bf q},-\omega) }
\right]^{\rm T}
\cdot {\cal Q}({\bf q},\omega) \cdot
\left[\matrix{
{\hat R}_\parallel({\bf q},\omega) \cr 
{\hat R}_\perp({\bf q},\omega) \cr  
R_\parallel({\bf q},\omega) \cr  
{1\over v} R_\perp({\bf q},\omega)  
}\right], 
\end{equation}
where 
\[
{\cal Q}({\bf q},\omega)=\left[\matrix{
-C_\parallel(\omega) & -{1\over 2}C_\times(\omega) &
K_{11}q^2-i\omega A_{11} & vK_{12}q^2+i\omega A_{12} \cr \cr
 -{1\over 2}C_\times(-\omega)   & -C_\perp(\omega) &
K_{21}q^2-i\omega A_{21} & vK_{22}q^2-i\omega A_{22} \cr \cr
K_{11}q^2+i\omega A_{11} & K_{21}q^2+i\omega A_{21} & 0 & 0 \cr \cr
vK_{12}q^2-i\omega A_{12} & vK_{22}q^2+i\omega A_{22} & 0 & 0 \cr
}\right].
\]

\begin{multicols}{2}
Neglecting all higher order terms 
in the action, we arrive at a Gaussian theory, in which different
Fourier modes are decoupled, and which can be solved by inverting 
the matrix in Eq.(\ref{qform}). (See Appendix \ref{invert}.) 
The quadratic action (\ref{quadratic}) remains invariant 
under the scale transformation
\begin{equation}
\label{Gaussscale}
\begin{array}{ll}
{\bf x}\rightarrow b{\bf x},\quad &t\rightarrow b^2 t, \cr \\
R_\parallel\rightarrow b^{2-d/2}{R_\parallel},\quad 
&R_\perp\rightarrow b^{2-d}{R_\perp}, \cr \\
{\hat R}_\parallel\rightarrow b^{-2-d/2}{{\hat R}_\parallel},\quad &
{\hat R}_\perp\rightarrow b^{-2-d/2}{{\hat R}_\perp}, \cr \\
v\rightarrow b^{-d/2}v,\quad &F-F_{\rm MF}\rightarrow b^{-d/2}(F-F_{\rm MF}), 
\end{array}
\end{equation}
except for terms proportional to $K_{12}$ and $K_{22}$ which vanish
at the depinning transition as $v\to 0^+$.
For $d>4$, all higher order terms in $\tilde S$ decay away upon rescaling,
and we recover an asymptotically quadratic theory with critical exponents
$\beta=1,\,z_\parallel=2,\,\nu=2/d,\,\zeta_\parallel=(4-d)/2,\,
\zeta_\perp=2-d$. The remaining exponent, $z_\perp$, 
can be found by comparing the static 
and dynamic parts of the transverse linear response. This gives 
$z_\perp=2+d/2=z_\parallel+1/\nu$, as shown previously by the
 exponent identity (\ref{exp6}).
The exponents related to longitudinal fluctuations, not surprisingly,
are identical to corresponding exponents in the interface problem\cite{NF93}.
However, we have also calculated new exponents characterizing transverse 
fluctuations. We see that even the simple Gaussian theory exhibits
anisotropic exponents.
 
At $d=d_c=4$ dimensions, the scaling dimension of $R_\parallel$ changes sign
and we cannot neglect its higher powers anymore. Simple dimensional 
analysis indicates that the only higher order terms in $\tilde S$ which become 
marginal at $d=d_c$ involve vertex functions 
$U_{\alpha\gamma,n}$, given in Eq.(\ref{m2series}). 
This series can be summed up over $n$, together with the $n=0$ term
$C_{\alpha\gamma}$ included in the Gaussian theory, 
to yield
\begin{eqnarray}
\frac{1}{2}\int&{d^d\!x\,}&{dt\,} dt'{\hat R}_\alpha({\bf x},t)
{\hat R}_\gamma({\bf x},t') \nonumber \\
& &\times C_{\alpha\gamma}\left(v(t-t')
+R_\parallel({\bf x},t)-R_\parallel({\bf x},t')\right).
\end{eqnarray}
All higher order terms in $\tilde S$ are formally irrelevant since
they involve additional powers of ${\hat R}_\parallel$, ${\hat R}_\perp$, or
$R_\perp$, whose scaling exponents are less than zero.  

For $d<d_c$, the vertex functions $U_{\alpha\gamma,n}$ become 
more and more relevant for increasing $n$ under the rescaling 
(\ref{Gaussscale}), 
and the fixed point moves away from the Gaussian theory. 
In $d=4-\epsilon$ dimensions, we look 
for new fixed points with different scaling properties:
\begin{equation}
\label{Newscale}
\begin{array}{ll}
{\bf x}\rightarrow b{\bf x},\quad &t\rightarrow b^{z_\parallel} t,\cr \\
R_\parallel\rightarrow b^{\zeta_\parallel}{R_\parallel},\quad &
R_\perp\rightarrow b^{\zeta_\perp}{R_\perp}, \cr \\
{\hat R}_\parallel\rightarrow b^{\theta_\parallel-d}{{\hat R}_\parallel}
,\quad &
{\hat R}_\perp\rightarrow b^{\theta_\perp-d}{{\hat R}_\perp}, \cr \\
F-F_{\rm MF}\rightarrow b^{-1/\nu}(F-F_{\rm MF}),\quad &{\bf v}\rightarrow 
b^{-\beta/\nu}{\bf v}.  
\end{array}
\end{equation}
To calculate the new exponents to first order
in $\epsilon$, we employ a one-loop momentum shell RG scheme,
treating all non-Gaussian terms in the action  (i.e. $\cal U$
in Eq.(\ref{m2series})), 
as a perturbation. Perturbative calculations proceed by expanding
$\langle e^{\cal U}\rangle_0$, where $\langle\cdots\rangle_0$ 
denotes averaging with respect to the  Gaussian action $\tilde S_0$, in
powers of ${\cal U}$. A renormalization transformation is then
constructed as follows: 
{\bf (1)} Perform the averages only over short wavelength fluctuations 
${\bf \hat R}^>,{\bf R}^>$ with wavenumbers $\Lambda/b<|{\bf q}|<\Lambda$, 
where $b=e^{\delta\ell}$. The resulting coarse grained action is
perturbatively given by
\begin{equation}
{\tilde S^<}={\tilde S_0^<}+\langle{\cal U}\rangle^>_0
+{1\over 2}\langle{\cal U}^2\rangle^>_{0,c}+ O({\cal U}^3).
\end{equation}
{\bf (2)} Apply the rescaling transformations given in 
(\ref{Newscale}), bringing back the short-distance cutoff $\Lambda$ 
to its original value. {\bf (3)} The exponents are then determined
from the fixed points associated with the RG flows of the
the action. Since Models A and B are characterized by distinct
fixed points, we shall discuss them separately.

\end{multicols}

\subsection{Model A}

In the low-frequency, small-wavevector limit, the effective action
for Model A is
\begin{eqnarray}
\label{actionA}
\tilde S^{(A)} &=& -\int{dt\,}{d^d\!x\,} [F-F_{\rm MF}(v)]
{\hat R}_\parallel({\bf x},t)  \\
& & \quad -\int\limits_{{\bf q},\omega}\left\{
{\hat R}_\parallel(-{\bf q},-\omega)R_\parallel({\bf q},\omega)
(Kq^2-i\omega A_{11})
+{\hat R}_\perp(-{\bf q},-\omega)R_\perp({\bf q},\omega)
(Kq^2-i\omega A_{22}/v)
\right\} \nonumber \\
& & \quad+\frac{1}{2}\int{d^d\!x\,}{dt\,}{dt\,}'
{\hat R}_\parallel({\bf x},t){\hat R}_\parallel({\bf x},t')
C_\parallel\left(v(t-t')+R_\parallel({\bf x},t)-R_\parallel({\bf x},t')\right) 
\nonumber \\
& & \quad+\frac{1}{2}\int{d^d\!x\,}{dt\,}{dt\,}' 
{\hat R}_\perp({\bf x},t){\hat R}_\perp({\bf x},t')
C_\perp\left(v(t-t')+R_\parallel({\bf x},t)-R_\parallel({\bf x},t')\right).
\nonumber
\end{eqnarray}

\begin{multicols}{2}

The Gaussian part has the correlation functions,
\begin{mathletters}
\label{corrA}
\begin{eqnarray}
\langle{\hat R}_\parallel(-{\bf q},-\omega)R_\parallel({\bf q},\omega)
\rangle_0 &=&
\frac{1}{Kq^2-i\omega A_{11}} ,\\
\langle{\hat R}_\perp(-{\bf q},-\omega)R_\perp({\bf q},\omega)\rangle_0 &=&
\frac{1}{Kq^2-i\omega A_{22}/v} ,\\
\label{Cpara}
\langle R_\parallel(-{\bf q},-\omega)R_\parallel({\bf q},\omega)\rangle_0 &=&
\frac{C_\parallel(\omega)}{K^2q^4+(\omega A_{11})^2}, \\
\label{Cperp}
\langle R_\perp(-{\bf q},-\omega)R_\perp({\bf q},\omega)\rangle_0 &=&
\frac{C_\perp(\omega)}{K^2q^4+(\omega A_{22}/v)^2}.
\end{eqnarray}
\end{mathletters}
The vertex functions $U_{\alpha\gamma,n}=0$ for $\alpha\neq
\gamma$, and these terms are not generated by the RG transformation.
The renormalization of remaining vertex functions $U_{\parallel,n}$,
and $U_{\perp,n}$ for $n>0$ can be recast into a functional 
renormalization of $C_{\parallel}(vt)$ and $C_{\perp}(vt)$, 
provided that $vt$ and $R_\parallel$ scale in the same way, i.e.
$\zeta_\parallel=z_\parallel-\beta/\nu$.
This relation can be independently obtained from Eqs.~(\ref{Aexp1}) 
and (\ref{Aexp2}), derived in 
Sec.~\ref{model} from more general (and nonperturbative) arguments.
The renormalized vertex functions are then obtained from
successive derivatives of $C(vt)$ as
\begin{equation}
U_{\alpha,n}(u)=C^{(n)}_{\alpha}(u).
\end{equation}
This ensures that the form of Eq.(\ref{actionA}) is retained under
renormalization, albeit with renormalized parameters. 
Eqs.~(\ref{Cpara}) and (\ref{Cperp}) suggest that 
$C_\alpha(vt)$ may be interpreted as {\it temporal}  
correlation functions of an effective force generated
by the quenched disorder. 

The renormalization of some terms in Eq.(\ref{actionA}) do not get any 
contribution from the momentum shell averaging step, giving rise to 
additional exponent relations that are correct to all orders in the 
$\epsilon$ expansion. 
The first relation is due to the fact that 
$F$ never appears explicitly 
in any of the contractions or higher order vertex functions. 
Thus, the renormalization of the term proportional to
$F-F_{{\rm MF}}$ can be written as
\begin{equation}
\frac{\partial(F-F_{{\rm MF}})}{\partial\ell}=
(z_\parallel+\theta_\parallel)(F-F_{{\rm MF}})+{\rm constant},
\end{equation}
where ``constant" refers to an expression that does not involve $F$.
This RG flow equation can be rewritten as 
\begin{equation}
\frac{\partial(F-F_c)}{\partial\ell}=
(z_\parallel+\theta_\parallel)(F-F_c),
\end{equation}
with a suitable choice of $F_c$. Hence, higher order 
corrections may shift the threshold force, but do not influence 
the scaling of $F-F_c$. This implies that
\begin{equation}
\label{exp7}
z_\parallel+\theta_\parallel-1/\nu=0.
\end{equation}
Furthermore, there are no contractions that contribute to the 
renormalization of $K$ or $A_{22}$. Thus,
\begin{eqnarray}
\label{exp8}
\theta_\parallel+z_\parallel+\zeta_\parallel-2 &=&  0, \\
\label{exp9}
\theta_\perp+\zeta_\perp+\beta/\nu &=&0 ,
\end{eqnarray}
respectively. 
As a result, all critical exponents are determined in terms of 
$\zeta_\parallel$, $\zeta_\perp$ and $z_\parallel$. 
These exponents can be computed by constructing RG flow equations
for the remaining parameters. 

\subsubsection{Renormalization of~~$C_{\alpha}$}

After performing the momentum shell integration and rescaling, 
details of which are given in Appendix~\ref{Capp}, we 
arrive at the recursion relations for the renormalized functions
$C_{\alpha}(u)$:
\begin{eqnarray}
\label{RRpara}
\frac{\partial C_\parallel(u)}{\partial\ell} &=&
[\epsilon+2\theta_\parallel+2(z_\parallel-2)]C_\parallel(u)
+\zeta_\parallel u C'_\parallel(u)  \\
&-&{K_d}\left\{[C'_\parallel(u)]^2 +
[C_\parallel(u)-C_\parallel(0)]
C''_\parallel(u)\right\}, \nonumber \\
\label{RRperp}
\frac{\partial C_\perp(u)}{\partial\ell} &=&
[\epsilon+2\theta_\perp+2(z_\parallel-2)]C_\perp(u)  \\
&+&\zeta_\parallel u\,C'_\perp(u) 
-{K_d}\left\{[C_\parallel(u)-C_\parallel(0)]
C''_\perp(u)\right\}. \nonumber
\end{eqnarray}
The constant $K_d\equiv S_d\Lambda^{d-4}/[(2\pi)^d K^2]$, 
where $S_d$ is the total solid angle in $d$-dimensions.
Primes denote derivatives with respect 
to $u$. Terms proportional to $uC'_\alpha(u)$ arise from the
rescaling of $u$. We look for fixed-point solutions $C^*_\alpha(u)$
that decay to 0 when $|u|$ is large, since they are related to
correlation functions of the system, which are expected to 
vanish for large time differences. 

Not surprisingly, the functional recursion relation for $C_\parallel(u)$
is identical to the one obtained in Ref.~\cite{NF93}. In fact, all
higher loop corrections are identical as well. This is in excellent
harmony with the argument presented in Sec.~\ref{model}, and allows
us to use the results of NF.
Setting $\partial C^*_\parallel/\partial\ell=0$, and integrating 
Eq.~(\ref{RRpara}) from $u=-\infty$ to $\infty$, we get
\begin{equation}
\label{intC}
[\epsilon+2\theta_\parallel+2(z_\parallel-2)-\zeta_\parallel]
\int\limits_{-\infty}^\infty C^*_\parallel(u)\,du = 0.
\end{equation}
Provided that the RG flows go to a fixed-point solution with
$\int C^*\neq 0$, this implies that $\zeta_\parallel=
\epsilon-2[2-(z_\parallel+\theta_\parallel)]$. The mean-field
correlation function satisfies this integral condition for both
random-field and random-bond disorder, since $C$ is essentially
insentitive to the value of the random potential between consecutive
local minima $P$, where the line moves quickly. There are other
fixed points with $\int C^*=0$, but they are irrelevant for our
discussion. 
Thus, from Eqs.~(\ref{Aexp1}) and (\ref{exp7}), we obtain
\begin{eqnarray}
\zeta_\parallel&=& \epsilon/3, \\
\nu &=& \frac{3}{6-\epsilon}.
\end{eqnarray}
NF prove that these results are correct to {\it all orders} in $\epsilon$,
by showing that the contributions to the 
renormalization of $C_\parallel(u)$ from higher-order 
terms is a complete derivative with respect to $u$.
Upon integration over $u$, such higher order terms do not
alter Eq.(\ref{intC}), leaving the exponents unchanged.  

Using $\zeta_\parallel=\epsilon/3$, an implicit solution for 
$C^*_\parallel(u)$ is obtained as
\[
C^*_\parallel(u)-C^*_\parallel(0)-C^*_\parallel(0)
\ln\left(\frac{C^*_\parallel(u)}{C^*_\parallel(0)}\right)
= {C^*(0)\over 2}\left(\frac{u}{u_0}\right)^2,
\]
where $u_0\equiv\sqrt{3K_dC^*(0)/\epsilon}$.
$C_\parallel^*(0)$ is arbitrary, and can be changed by a
rescaling of the fields $R_\parallel$. Expanding the logarithm 
for small $u$, we see that there is a kink at 
the origin, as
\begin{equation}
\label{Corigin}
\frac{C^*_\parallel(u)}{C^*_\parallel(0)}=\left[1-\frac{|u|}{u_0}
+\frac{1}{3}\left(\frac{u}{u_0}\right)^2\right]
+O(|u/u_0|^3).
\end{equation}
For  $|u|\gg u_0$, the fixed point solution behaves like a Gaussian, and
\[
C^*_\parallel(u)\approx C_\parallel^*(0) 
\exp\left[-{u^2\over 2u_0^2}\right]. 
\]

\begin{figure}
\narrowtext
\epsfxsize=2.9truein
\centerline{\epsffile{\dir/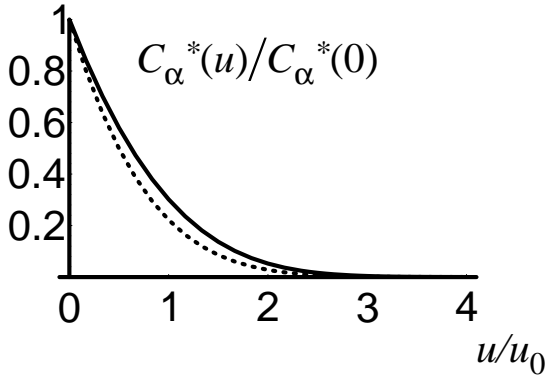}}
\medskip
\caption{Fixed point functions $C^*_\parallel(u)$ (solid line) and 
$C^*_\perp(u)$ (dotted line), normalized to yield 1 at the origin.
Their values for $u<0$ (not shown) are  
found from $C_\alpha^*(u)=C_\alpha^*(-u)$.}
\label{fig_Cfixed}
\end{figure}

We next examine the fixed-point solution $C^*_\perp(u)$, which is 
the new element of our computation. Setting  
$\partial C^*_\perp/\partial\ell=0$ 
and looking at the limit $u\to 0^+$, we get (assuming that 
$C^*_\perp(0^+)\ne 0$)
\begin{equation}
[\epsilon+2\theta_\perp+2(z_\parallel-2)]C^*_\perp(0^+)=0.
\end{equation}
Combined with Eqs.(\ref{Aexp1}),
(\ref{Aexp2}), and (\ref{exp9}), this result implies
\begin{equation}
\label{zetaperp}
\zeta_\perp=\zeta_\parallel-\frac{d}{2}=-2+\frac{5\epsilon}{6}.
\end{equation}
In Appendix~\ref{Cperpapp}, we show that this result is in fact correct
to all orders in $\epsilon$ since there are no contributions to 
$C^*_\perp(0^+)$ from momentum-shell integration.
The fixed point solution (for $u>0$) satisfies the equation
\begin{equation}
\label{Cfixed}
\frac{d}{du}\ln|C_\perp^{*'}(u)|=
\frac{u}{u_0^2}\left[
\frac{C^*_\parallel(u)}{C^*_\parallel(0)}-1\right]^{-1}.
\end{equation}
Upon integrating twice, Eq.~(\ref{Cfixed}) leads to
\end{multicols}
\begin{equation}
C^*_\perp(u)= -{C^{*'}_\perp(0^+)}
\int\limits_u^\infty du'\,\exp\left\{-\frac{1}{u_0^2}
\int\limits_{0^+}^{u'}du''\,\frac{u''}{1-
\left[C^*_\parallel(u'')/C^*_\parallel(0)\right]}\right\},
\end{equation} 
\begin{multicols}{2}
where $C^{*'}_\perp(0^+)$ is arbitrary in the same sense 
as $C^*_\parallel(0)$. For $|u|\gg u_0$, Eq.(\ref{Cfixed})
gives
\begin{equation}
C^*_\perp(u) \approx C \frac{u_0^2}{u}\exp\left[-{u^2\over 2u_0^2}\right],
\end{equation}
where $C$ is a constant related to 
$C_\perp^{*'}(0^+)$. 
The numerical solution for the fixed point functions $C_\alpha^*(u)$
are shown in Fig.~\ref{fig_Cfixed}. 
The qualitative features of $C^*_\parallel$ and $C^*_\perp$ are 
similar: both have a discontinuous derivative at the origin, 
and decay as a Gaussian for large
values of $|u|$. However, note that their scaling dimensions 
differ by $\zeta_\parallel$. 

The exponent $\zeta_\parallel=\epsilon/3$ can also be obtained by
naive dimensional arguments: In dimensions $d>4$, the random force
can be expanded as ${f_\parallel}({\bf x},{r_\parallel},{r_\perp})
={f_\parallel}({\bf x},0,0)+O({r_\parallel},{r_\perp})$.
Since both ${r_\parallel}$ and ${r_\perp}$ have negative scaling dimensions 
$(\zeta_\parallel,\zeta_\perp<0)$, the correction terms can be ignored.
The random force scales as $b^{-d/2}$ under a scaling ${\bf x}\to b{\bf x}$,
leading to the Gaussian roughness of $\zeta_\parallel=2-d/2$. A 
similar scaling argument applied to Eq.(\ref{Astar}) leads to
$\zeta_\perp=\zeta_\parallel-d/2=2-d$. For $d<4$, the scaling
dimension of ${r_\parallel}$ is positive, and higher powers of 
${r_\parallel}$ in an
expansion of ${f_\parallel}({\bf x},{r_\parallel},{r_\perp})$ are more 
relevant. It is then reasonable
to assume that in this case the statistical properties of ${f_\parallel}$ at
large ${r_\parallel}$ are crucial. If uncorrelated at large separation,
${f_\parallel}({\bf x},{r_\parallel},0)$ scales as 
$b^{-(d+\zeta_\parallel)/2}$. When 
equated to $b^{\zeta_\parallel-2}$ for the scaling of 
$\nabla_{\bf x}^2{r_\parallel}$,
this leads to $\zeta_\parallel=\epsilon/3$ in agreement with the RG
treatment. Essentially, the statement regarding the non-renormalization of
$\int du\,  C_\parallel(u)$ justifies the above ``naive" scaling.
However, a similar reasoning from Eq.(\ref{Astar}) would have concluded
$\zeta_\perp-\zeta_\parallel=-(d+\zeta_\parallel)/2$, in disagreement
with Eq.(\ref{zetaperp}). In this case, $\int du\,  C_\perp(u)$ {\it is}
renormalized, but $C_\perp(0)$ is not; suggesting that despite the 
presence of relevant higher order powers in the expansion of 
${f_\perp}({\bf x},{r_\parallel},{r_\perp})$ around ${\bf r}={\bf 0}$, 
the scaling properties are 
still controlled by ${f_\perp}({\bf x},0,0)$. We have no physical motivation
for this rather curious conclusion. 

\subsubsection{Propagator Renormalization}
\label{propagator}

The only one remaining exponent is $z_\parallel$, which 
can be obtained by examining the renormalization of
$A_{11}$. One-loop contributions arise from the $n=2$ term in
$\langle{\cal U}\rangle_0^>$, which is
\begin{eqnarray*}
\frac{1}{4}\int d^dx\int\limits_{-\infty}^{\infty}&dt& 
\int\limits_{-\infty}^{\infty}dt' 
{\hat R}_\parallel({\bf x},t){\hat R}_\parallel({\bf x},t') \\
& &\times[R_\parallel({\bf x},t)-R_\parallel({\bf x},t')]^2 
C''_\parallel\left(v(t-t')\right).
\end{eqnarray*}
Replacing $[R_\parallel({\bf x},t')]^2$ with $[R_\parallel({\bf x},t)]^2$ 
does not change the integral. Thus, upon further manipulation, this
term in the action can be written as 
\begin{eqnarray*}
\int d^dx\int\limits_{-\infty}^{\infty}&dt&\int\limits_{-\infty}^{t}dt'
{\hat R}_\parallel({\bf x},t){\hat R}_\parallel({\bf x},t')
R_\parallel({\bf x},t) \\
& &\times[R_\parallel({\bf x},t)-R_\parallel({\bf x},t')]
C''_\parallel\left(v(t-t')\right) .
\end{eqnarray*}
Since a contraction forces $t$ and $t'$ to be within $O(1)$ of
each other, and we are only interested in the first time derivative, we 
can substitute $R_\parallel({\bf x},t)-R_\parallel({\bf x},t')\approx
(t-t')\partial_t R_\parallel({\bf x},t)$. Now, contracting 
${\hat R}_\parallel({\bf x},t')$ with $R_\parallel({\bf x},t)$ and integrating
over the momentum shell, we obtain a contribution to $A_{11}$ equal to
\begin{equation}
-\delta\ell \frac{S_d\Lambda^{d}}{(2\pi)^{d}A_{11}}
\int\limits_0^\infty d\tilde t\,\tilde t\,
e^{-K\Lambda^2\tilde t/A_{11}}C''_\parallel(v\tilde t\,).
\end{equation} 
The minus sign comes from the opposite overall signs of $m=1$ and 
$m=2$ terms in Eq.~(\ref{actionA}).
For $v\to 0$, we can set the argument of $C''_\parallel$ to zero. 
However, this causes a problem: $C''_\parallel$ has a term proportional
to $\delta(vt)$ in the low-frequency analysis, this term diverges as 
$1/v$ for $vt\to 0$. This apparent divergence cannot be avoided within the
low-frequency analysis we have used so far. The propagator is 
sensitive to high-frequency behavior of the vertex functions.
Careful analysis of the high frequency structure of $C''_\parallel$
shows that the terms that contribute to the diverging part of 
$C''_\parallel(0)$ {\it do not} enter the renormalization of the
propagator. (See Appendix~\ref{hifreq}.) This is essentially due to 
the causal nature of the response: Perturbations right after a 
jump do not influence the motion before the jump.  
The correct way to avoid
these divergent terms within the low-frequency analysis is to use
$C''_\parallel(0^+)$ instead of $C''_\parallel(0)$. Near the fixed
point, this can be 
calculated to $O(\epsilon)$ from Eq.~(\ref{Corigin}) as 
$C^{*''}_\parallel(0^+)=2\epsilon/(9K_d)$, resulting in
\[
A_{11}^<=A_{11}-\delta\ell A_{11} K_d C^{*''}_\parallel(0^+) 
= A_{11}[1-\delta\ell(2\epsilon/9)].
\]
Finally, after performing the rescaling, we obtain the recursion relation
\begin{equation}
\frac{\partial A_{11}}{\partial \ell}=A_{11}[\theta_\parallel+\zeta_\parallel
-2\epsilon/9],
\end{equation}
which yields
\begin{equation}
z_\parallel=2-2\epsilon/9+O(\epsilon^2).
\end{equation}

\subsection{Model B}

The presence of  off-diagonal terms in the action changes the critical
scaling properties of Model B. The nonzero contractions that appear
in the momentum shell integration in this case are (cf. Appendix~\ref{invert})
\begin{mathletters}
\label{corrB} 
\begin{eqnarray}
\langle{\hat R}_\parallel(-{\bf q},-\omega)R_\parallel({\bf q},\omega)
\rangle_0 &=&
\frac{1}{K_\parallel q^2-i\omega\rho_\parallel} ,\\
\langle{\hat R}_\perp(-{\bf q},-\omega)R_\parallel({\bf q},\omega)\rangle_0 
&=&
\frac{\kappa}{K_\parallel q^2-i\omega\rho_\parallel} ,\\
\label{Rpara}
\langle R_\parallel(-{\bf q},-\omega)R_\parallel({\bf q},\omega)\rangle_0 &=&
\frac{{\tilde C}(\omega)}{K_\parallel^2q^4+\omega^2\rho_\parallel^2}, 
\end{eqnarray}
\end{mathletters}
where 
\begin{eqnarray*}
\kappa &\equiv& A_{12}/A_{22}, \\
K_\parallel &\equiv& K_{11}+\kappa K_{21}, \\
\rho_\parallel &\equiv& A_{11}+\kappa A_{21}, \\ 
{\tilde C}(\omega) &\equiv& C_\parallel(\omega)
+\kappa {\rm Re}[C_\times(\omega)]+\kappa^2 C_\perp(\omega). 
\end{eqnarray*}

In addition to the nonrenormalization relations (\ref{exp7}-\ref{exp9}),
the nonrenormalization of $K_{21}$ or $A_{12}$ dictates that
\begin{equation}
\theta_\parallel=\theta_\perp.
\end{equation}
This immediately implies the exponent identity
\begin{equation}
\zeta_\perp=\zeta_\parallel-1/\nu.
\end{equation} 
The naive scaling argument based on Eq.(\ref{Astar}) gives an
equivalent result when the scaling dimension of 
$\partial{r_\perp}/\partial{r_\parallel}$
($\zeta_\perp-\zeta_\parallel$) is equated to the scaling dimension
of ${f_\perp}({\bf x}, {r_\parallel}, 0)$ [$-(d+\zeta_\parallel)/2$]. 
The naive argument 
works this time, since $\int du\,C_\perp(u)$ remains finite at the 
fixed point (see below).

Under this rescaling, $\kappa$ and $K_\parallel$ remain unrenormalized, and 
the renormalizations of $\rho_\parallel$ and $\tilde C$ determine the
remaining exponents $\zeta_\parallel$ and $z_\parallel$.
The recursion relations of vertex functions 
$C_{\alpha}$ are more complicated, but there is a relatively
simple fixed point solution with
\begin{equation}
\label{ctilde}
\tilde C^*(u)=4C^*_\parallel(u)=2\kappa C^*_\times(u)
=4\kappa^2C^*_\perp(u).
\end{equation}
Furthermore, $\tilde C(u)$ satisfies a recursion
relation identical to that of $C_\parallel(u)$ given in Eq.(\ref{RRpara}).
This result once more shows that longitudinal fluctuations, whose
correlations are given by Eq.(\ref{Rpara}), are not altered by
the introduction of transverse fluctuations even in the more
general case of Model B. 

The renormalization of $\rho_\parallel$ also
gives results very similar to that of Model A, with the substitutions 
$C_\parallel''\to{\tilde C}''$ and $A_{11}\to\rho_\parallel$. Thus, the RG
analysis gives the same exponents $\zeta_\parallel=\epsilon/3$
and $z_\parallel=2-2\epsilon/9+O(\epsilon^2)$. Further details
appear in Appendix~\ref{modelB}.

If the Hall Angle $\phi$ is sufficiently small, the FL can not
distinguish between zero and nonzero angles. Therefore, the
effective roughness and dynamic exponents at small length and
time scales should be given by the Model A fixed point.
Note that $\kappa=\tan\phi$ in an isotropic
system with nonzero Hall angle (cf. Eq.(\ref{dFdv})), and $\kappa$ is
in general strongly related to the macroscopic Hall angle. Thus,
$\kappa \ll 1 $  when the system is almost Model A-like, and
its nonrenormalization determines the cross-over behavior to
the Model B fixed point: Under renormalization with Model A exponents, 
the system remains near the Model A fixed point until the ratio 
$C_\perp/C_\parallel$ increases to $O(\kappa^{-2})$, as the
Model B fixed point is approached. Isotropic effective exponents 
appear in this crossover regime.
The length scale $\xi_\times$ at which the behavior crosses over 
to the Model B is roughly given by
\[
\phi\approx\xi_\times^{\zeta_\perp-\zeta_\parallel},
\]
(with Model A exponents for $\zeta_\alpha$), 
i.e. the anisotropy is noticeable when the angular spread in the
direction of a typical avalanche is of the order of $\phi$.
Thus, for the FL,
\[
\xi_\times\sim\phi^{-2},
\] 
which diverges as $\phi\to0$. When $\xi<\xi_\times$, the anisotropic
fixed point is never approached. Thus, the true critical region
can be very small and difficult to observe for small Hall angle.

\section{Numerical Work}
\label{numerical}

In this section, we present and discuss the results obtained by
numerically integrating Eqs.~(\ref{motion}), providing a test
of the analytical results presented so far. 
There are several difficulties associated with numerically studying
critical behavior in a finite system slightly above threshold. 
In order to obtain meaningful statistical averages
one must wait for the system to reach a stationary state. However,
for any reasonably broad distribution of pinning forces, the system
always gets pinned after a time $\sim e^{(F-F_c)^\nu L}$, where $L$
is the linear extension. Therefore, in order to probe the
critical region, it is necessary to go to very large system sizes.

The necessity of integrating big systems, and the 
large computational cost of implementing quenched
disorder, forced us to restrict numerical simulations to $d=1$,
in any case the most physically relevant dimensionality. 
We were further motivated by the expectation that some
exponents were calculated to all orders in $\epsilon$, and thus could
be checked even at $\epsilon=3$.

Integrations were carried out as follows: Coordinates $x$ and $t$
were discretized, but the position ${\bf r}$ was left continuous.
For each $x$, the value of the random potential at point
${\bf r}$ was determined from a superposition of arttactive impurity 
potentials
\[
U_i({\bf r}')={1\over 2}s_i(r'^2-r_0^2)\Theta(r_0-r'),
\] 
where $\Theta$ is the step function and $r'$ is the distance from the
center of the impurity. The impurity centers 
were randomly placed with a density $w$; their strengths $s_i$
were randomly drawn from a uniform distribution
$[0,s_{\rm max})$. The range  $r_0$, of the impurity potential was kept 
constant. This construction creates a random scalloped potential landscape,
eliminating any additional crossover effects that could arise
from a smooth potential.

Unless noted otherwise, all presented results were obtained
using a grid size $\Delta x=1$, and a time step $\Delta t=0.02$, 
in order to optimize
computational constraints. (Smaller values of $\Delta x$ or $\Delta t$
did not lead to significant improvements.) 
Free boundary conditions were preferred over periodic ones
since scaling was observed over a larger range of length scales
in the former case.
Other simulation parameters
were $K=1$, $w=1$, $r_0=1$, $s_{\rm max}=2$.  We expected 
a threshold force close to $1$ for these parameters. A summary of
our findings is presented below. 

\begin{figure}
\narrowtext
\epsfxsize=3.2truein
\epsffile{\dir/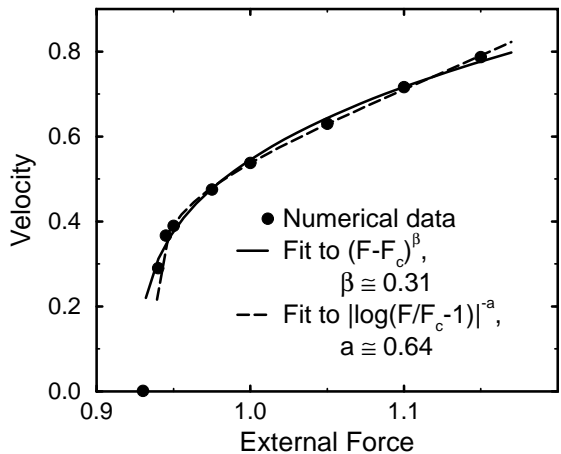}
\caption{A plot of average velocity versus external force for 
a system of size 2048. Statistical errors are smaller than 
symbol sizes. Both fits have three adjustable parameters: The
threshold force, the exponent, and an overall multiplicative constant.}
\label{vfplot}
\end{figure}

The velocity exponent $\beta$ can be extracted from a plot 
of velocity versus external force. Such a plot is given in 
Fig.~\ref{vfplot} for a system of size $L=2048$. Each data point
was obtained by a time average over $10^5$ time units and took
about 30 hours of CPU time on a Silicon Graphics R4000 workstation. 
The best power law fit gives an exponent $\beta\approx 0.31$, but a 
weaker logarithmic dependence, which corresponds to $\beta=0$, 
seems to provide a better fit to the data. The conclusion is that
higher order terms in $v$ give very large corrections to the scaling
of $v$, since either $\beta$ is very small or exactly zero.
$\beta=0$ would imply that $z_\parallel=1$, a possibility discussed 
by NF for interfaces in $1+1$ dimensions\cite{NF93}.
The threshold force $F_c$, is between 0.93 and 0.94.

The roughness exponents $\zeta_\parallel,\zeta_\perp$ are extracted
from equal-time correlation functions
\[
\left<[r_\alpha({\bf x},t)-r_\alpha({\bf x}',t)]^2\right>
\sim |{\bf x}-{\bf x}'|^{2\zeta_\alpha}.
\]

\begin{figure}
\epsfxsize=3.2truein
\epsffile{\dir/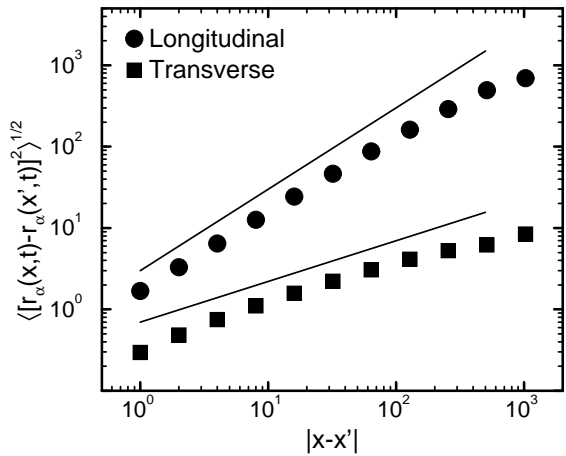}
\caption{A plot of equal time correlation functions versus separation,
for a system of size 2048 at $F=0.95$. The observed roughness
exponents are close to the theoretical predictions of 
$\zeta_\parallel=1,\;\zeta_\perp=0.5$, which are shown as solid lines
for comparison.}
\label{xcorr}
\end{figure}

Results for a system of size 2048, at a driving force 
of 0.95 [$(F-F_c)/F_c\approx 10^{-2}$], 
are shown in Fig.~\ref{xcorr}. The averages were
taken over a time interval of $10^5$, after waiting for all
correlations to reach steady-state. The results are in overall
agreement with the predicted values of the exponents, even at 
$\epsilon=3$. The slightly smaller value of $\zeta_\parallel$ is expected,
since determination of the roughness exponent from equal-time
correlations becomes unreliable as the exponent approaches unity,
and is inappropriate if  it exceeds 1\cite{rbigexp}. 
The deviations
of transverse correlations from the scaling form are likely to
be due to crossover effects: The analysis of transverse
fluctuations in the critical region is correct only when $v/F \ll 1$,
because then the static part of the transverse propagator can be
neglected. However, in our simulations $v/F\approx 0.4$, suggesting that
the critical region is much smaller for transverse fluctuations
compared to longitudinal ones.

In order to obtain an independent estimate of the dynamical exponent 
$z_\parallel$, we also examined fluctuations in the spatially averaged
velocity as a function of time. The resulting measurements were related 
to the previously defined exponents by the following argument\cite{Parisi}:
Slightly above threshold, the motion of the line can be thought as
a superposition of avalanches of various sizes $l$, with an average
lifetime $l^{z_\parallel}$ and moment $l^{d+\zeta_\parallel}$.
Such avalanches occur if a portion of the line finds a region 
of size $l^{d+\zeta_\parallel}$ with weak impurities.
Thus, ignoring all power-law prefactors, the probability of 
such an avalanche for $l\gg\xi$ is 
\[
P(l)\sim\exp\left\{-\left({l/\xi}\right)^{d+\zeta_\parallel}\right\}.
\]
Velocities at two separate times are correlated if there is
an avalanche that is active at both times. Therefore, it is
reasonable to assume that at large times,
the contribution of an avalanche of size $l$ to $\left<v(t)v(0)\right>_c$ 
is proportional to $e^{-t/l^{z_\parallel}}$, once again neglecting
power-law prefactors that depend, for example, on the typical number
of active sites at a given time during the avalanche. The total 
contribution of all avalanches is given by an integral over all 
sizes $l$ with the probability measure $P(l)\,dl$. The leading-order
time dependence of the exponent can be determined by a saddle point 
evaluation of the integral, resulting in
\[
C_{\rm v}(t)=\frac{\left<v(t)v(0)\right>_c}{\left<v^2
\right>_c} \sim e^{-(t/\tau)^\gamma},
\]

\begin{figure}
\narrowtext
\epsfxsize=3.2truein
\epsffile{\dir/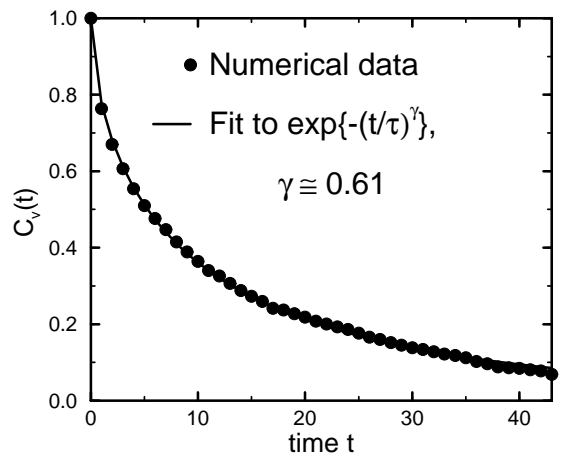}
\caption{Velocity correlations versus time, for the same system 
in Fig.~\protect\ref{xcorr}. A stretched exponential is a good fit
to the data.}
\label{vcorr}
\end{figure}

\noindent where $\gamma=(d+\zeta_\parallel)/(z_\parallel+d+\zeta_\parallel)$,
suggesting a stretched exponential.
The numerical results and the fit to a stretched exponential are shown 
in Fig.~\ref{vcorr}. (It should be noted that a comparable fit can 
also be achieved by a sum of two exponentials.)
Assuming that $\zeta_\parallel=1$, we arrive 
at $z\approx 1.3$, which is consistent with the value of 
$\beta\approx 0.31$ found from the velocity-force relation.
Unfortunately, the data becomes noisy at larger values of $t$, due to
the finite size of the time window used to extract the correlation 
function. The small value of $\tau(\approx 10)$ makes it hard to
predict the reliability of this estimate, since the power-law prefactors 
may be large and nonnegligible for such moderate values of $t$. 
Unfortunately, improving on this simple estimate is difficult as the
determination of power-law prefactors requires a number
of additional assumptions that are hard to test.
Nevertheless, based on the accumulated numerical evidence it can be 
reasonably argued that $z_\parallel$ is between 1 and 
4/3, the $O(\epsilon)$ RG prediction.

Computed longitudinal exponents are also in good
agreement with results from 1+1 dimensional interface 
depinning models. Numerical integration of 
Eq.(\ref{eqA}a) 
for an elastic interface\cite{Dong} (no transverse component)
has yielded critical exponents $\zeta=0.97\pm0.05$ 
and $\nu=1.05\pm0.1$. Similarly, the force vs. velocity
data has been adequately described by both a velocity 
exponent $\beta=0.24\pm0.1$ and  
a logarithmic dependence $v\sim1/\ln(F-F_c)$, which
corresponds to $\beta=0$. These results provide strong 
support for our prediction that longitudinal exponents are
unchanged when transverse fluctuations are introduced.
However, it should also be noted that experiments and 
various discrete models of interface growth 
have resulted in scaling behaviors that differ from 
system to system. A number of different experiments 
on fluid invasion in porous media\cite{rIntexp} 
give roughness exponents of around 0.8,
while imbibition experiments\cite{rBul,eFCA} have resulted 
in $\zeta\approx0.6$. Some of these results can be explained
by the effect of anisotropy, which will be discussed in 
the next Section. On the other hand, a discrete model studied 
by Leschhorn\cite{rHeiko}  
gives a roughness exponent of 1.25 at threshold.
Since the expansion leading to Eqs.(\ref{motion})
breaks down when $\zeta$ approaches one, it is not clear 
how to reconcile the results of Leschhhorn's numerical 
work\cite{rHeiko} with the coarse-grained description of 
the RG calculation, especially since any model with 
$\zeta>1$ cannot have a coarse grained description 
based on gradient expansions. 

\section{Discussion and Conclusions}
\label{conclusion}

In order to put the results we have found so far in better perspective, 
it is useful to discuss the effect of nonlinear terms that were 
ignored earlier, aspects of universality, and possible generalizations 
to other systems. These issues are discussed below.

\subsection{Nonlinear Terms}
The leading order nonlinearities in Eq.(\ref{nleqn}) can be
examined by a gradient expansion,  being careful 
to treat terms of $O((\partial_x r)^2, (\partial_x r)^2\partial_t r)$
accurately. After some rearrangement, we arrive at

\begin{mathletters}
\label{fulleqn}
\begin{eqnarray}
\frac{\eta\partial_t{r_\parallel}}{\sqrt{1+{s_\parallel}^2}}&=&
K_{11}\partial_x^2{r_\parallel}+K_{12}\partial_x^2{r_\perp} 
+ {\lambda_{1\parallel}\over 2}{s_\parallel}^2
+ {\lambda_{1\perp}\over 2}{s_\perp}^2 \nonumber \\
& & \;
+ \lambda_{1\times}{s_\parallel}{s_\perp}
+F+\tilde{f_\parallel}(x,{\bf r}, {s_\parallel}, {s_\perp}),  \\
\frac{\eta\partial_t{r_\perp}}{\sqrt{1+{s_\perp}^2}}&=&
K_{21}\partial_x^2{r_\parallel}+K_{22}\partial_x^2{r_\perp} 
+ {\lambda_{2\parallel}\over 2}{s_\parallel}^2
+ {\lambda_{2\perp}\over 2}{s_\perp}^2 \nonumber \\
& & \;
+ \lambda_{2\times}{s_\parallel}{s_\perp}
+\tilde{f_\perp}(x,{\bf r}, {s_\parallel}, {s_\perp}), 
\end{eqnarray}
\end{mathletters}
where ${s_\parallel}\equiv\partial_x{r_\parallel}, 
{s_\perp}\equiv\partial_x{r_\perp}$, 
and the random forces are
\begin{eqnarray*}
\tilde{f_\parallel}&=&\frac{({f_\parallel}-{s_\parallel} 
f_x)}{\sqrt{1+{s_\parallel}^2}}\cos\phi\nonumber \\
& & \;
+\frac{{s_\parallel}{s_\perp}{f_\parallel}
-[1+{s_\parallel}^2/2-{s_\perp}^2/2]{f_\perp}+{s_\perp} f_x}{\sqrt{
1+{s_\parallel}^2}}\sin\phi,  \\ 
\tilde{f_\perp}&=&\frac{({f_\perp}-{s_\perp} f_x)}{\sqrt{
1+{s_\perp}^2}}\cos\phi\nonumber \\
& & \;
+\frac{[1-{s_\parallel}^2/2+{s_\perp}^2/2]{f_\parallel}
-{s_\parallel}{s_\perp}{f_\perp}-{s_\parallel} f_x}{\sqrt{1+{s_\perp}^2}}
\sin\phi. 
\end{eqnarray*}
The remaining parameters are given by
\begin{eqnarray*}
F&=& \Phi_0 J, \\
\lambda_{1\parallel} &=&  -F\sin^2\phi,  \\
\lambda_{1\perp} &=&  -F\cos^2\phi,  \\
\lambda_{1\times} &=& -F\sin^2\phi, \\
\lambda_{2\parallel} &=&  F\sin\phi\cos\phi,  \\
\lambda_{2\perp} &=&  -F\sin\phi\cos\phi,  \\
\lambda_{2\times} &=& F\cos^2\phi.
\end{eqnarray*}
These equations of motion, and their generalizations to ${\bf x}\in\Re^d$,  
have thus been complicated by two factors: 
There are orientation-dependent terms, and the mean square of the random 
forces $\tilde\Delta_\alpha\equiv\langle\tilde{f_\alpha}^2\rangle$
also depend on the local orientation of the FL.
By naive dimensional counting, it can be immediately seen that
$\lambda_{1\parallel}$ and $\lambda_{2\parallel}$ 
are relevant with respect to the fixed points we have discussed 
for $d<4$.
In the case of Model A (isotropic disorder with $\phi=0$), 
Eq.(\ref{fulleqn}) further simplifies to
\begin{mathletters}
\label{isonlmotion}
\begin{eqnarray}
{\eta\partial_t{r_\parallel}\over\sqrt{1+{s_\parallel}^2}}&=&
K\partial_x^2{r_\parallel}-
\frac{F}{2}{s_\perp}^2+F 
+\frac{{f_\parallel}-f_x{s_\parallel}}{\sqrt{1+{s_\parallel}^2}}, \\
{\eta\partial_t{r_\perp}\over\sqrt{1+{s_\perp}^2}}&=&
K\partial_x^2{r_\perp}+F{s_\parallel}{s_\perp}+\frac{{f_\perp}
-f_x{s_\perp}}{\sqrt{1+{s_\perp}^2}}. 
\end{eqnarray}
\end{mathletters}
Note that the two relevant nonlinearities vanish,  
and that $\tilde \Delta_\alpha$
does not depend on orientation up to and including $O(s^2)$.  
Dimensional counting suggests that the remaining nonlinear terms
are irrelevant and Model A exponents are valid
for $d>1$. Many more nonlinear
terms become marginal at $d=1$, and the gradient expansion breaks down.
It is unlikely for the critical exponents to change their
value discontinuously at $d=1$, although logarithmic corrections
to scaling exponents are quite possible. 

The fixed point investigated here is unstable and 
only approached at the depinning force. Away from the threshold,
critical scaling laws are observed at scales smaller than the
correlation length scale $\xi$. Beyond this critical regime, the behavior of 
Eq.~(\ref{fulleqn}) is similar to regular diffusion with 
white noise (a multicomponent Edwards-Wilkinson (EW) equation\cite{EW}), 
or the generalized KPZ equation\cite{EKlines,EKpoly,Barabasi}). A nonzero 
$\lambda_{1\parallel}$ of $O(v)$ is generated kinetically in 
this regime even if the system is initially isotropic with $\phi=0$, 
due to the terms on the left-hand side of Eq.(\ref{fulleqn}a). 
For $d\leq2$, this nonlinearity is relevant, while for
$d>2$, a critical value $\lambda_c$ separates a  
weak-coupling region described by the EW equation from a strong-coupling
region described by the (generalized) KPZ equation
\cite{EKlines,EKpoly,Barabasi}.

When $\phi\neq0$, even in a fully isotropic medium, the relevant 
nonlinearities are nonzero, and the system is driven away from 
the ``linear" fixed points. We discuss this and other possibilities 
next.

\subsection{Anisotropy and Universality}
We noted earlier that anisotropy plays an important role in determining
scaling properties near depinning, even in the 
absence of nonlinear terms. To fully understand the effects of
anisotropy, {\it  including nonlinear terms}, let us start by considering 
the simplest prototype of a FL oriented 
along the $c-$axis of a High $T_c$ superconducting single crystal, 
such as YBCO. For simplicity, assume that the system is 
completely isotropic in the $y-z$ plane, with $\phi=0$.
Then, the motion of the FL is governed by Eqs.(\ref{isonlmotion}), 
and the only important source of anisotropy is due to 
$\langle {f_\parallel}^2\rangle=\langle {f_\perp}^2\rangle\neq\langle 
f_x^2\rangle$.
This causes the mean square magnitude of 
$\tilde{f_\parallel}$ to depend on the local orientation as,
\[
\tilde\Delta_\parallel\approx\langle{f_\parallel}^2\rangle +  
\left(\langle f_x^2\rangle-\langle{f_\parallel}^2\rangle\right)
{s_\parallel}^2, 
\]
For interfaces, the depinning force is known to scale with the
strength of the disorder\cite{Bruinsma,NSTL},  i.e.
$\tilde F_c\sim\Delta^{2/(4-d)}$. Thus, $\tilde\Delta_\parallel$
creates an orientation dependent depinning force\cite{TKD},
\begin{equation}
F_c({s_\parallel})\sim{\tilde\Delta_\parallel}^{2/(4-d)}
\sim F_c\left(1+\frac{2}{4-d}
\frac{\langle f_x^2\rangle-\langle{f_\parallel}^2\rangle}
{\langle{f_\parallel}^2\rangle}{s_\parallel}^2\right).
\end{equation}
This leads to a nonzero $\lambda_{1\parallel}$ when the nonlinear
corrections in Eq.(\ref{isonlmotion}) are taken into account.
For interfaces, the depinning transition with a nonzero 
$\lambda_{1\parallel}$ is thought to be equivalent to 
directed percolation depinning\cite{TKD}. Assuming that 
transverse fluctuations still do not affect longitudinal
ones, for $d=1$ the critical exponents $\zeta_\parallel$ and 
$\nu$ are related to the correlation length exponents 
$\nu_\parallel^{(DP)}$ and
$\nu_\perp^{(DP)}$ of directed percolation through
$\nu=\nu_\parallel^{(DP)}\approx1.73$ and
$\zeta_\parallel=\nu_\perp^{(DP)}/\nu_\parallel^{(DP)}\approx0.63$, 
while the dynamical exponent is $z_\parallel=1$. This in turn gives
$\beta=(z_\parallel-\zeta_\parallel)\nu
=\nu_\perp^{(DP)}-\nu_\parallel^{(DP)}\approx0.64$.

Using the connection to interface depinning further, 
we next consider tilting the FL away from the symmetry
axis $c$. In this case,  $\langle f_x {f_\parallel}\rangle$ and
$\langle f_x {f_\perp}\rangle$ are nonzero, and $F_c$ depends
linearly on $s_\parallel$, leading to terms proportional
to $\partial_x{r_\parallel}$ in the equation of motion. These further 
suppress the roughness exponent to $\zeta_\parallel=1/2$\cite{TKD}.
The analysis of transverse fluctuations for these two situations,
and many other possible ones, are complicated by the absence of
a suitable perturbative treatment. Different types of anisotropy
may lead to distinct transverse exponents even while the 
longitudinal ones remain identical. (Similar to the difference
between Models A and B,  although the latter is unstable to the
inclusion of nonlinear terms.) To systematically search
for universality classes, we may start with the most general equation of 
motion,  which has the gradient expansion,
\begin{eqnarray}
\label{genmotion}
\partial_tr_\alpha&=&\mu_{\alpha\beta}F_\beta
+\kappa_{\alpha\beta}\partial_xr_\beta
+K_{\alpha\beta}\partial_x^2r_\beta  \\
& &+\frac{1}{2}\lambda_{\alpha,\beta\gamma}
\partial_xr_\beta\partial_xr_\gamma
+\tilde f_\alpha(x,{\bf r},\partial_x{\bf r},\cdots)+\cdots, \nonumber
\end{eqnarray}
and with force-force correlations that depend on $\partial_x{\bf r}$.
Depending on the presence or absence of various terms allowed by 
symmetries, these equations encompass many distinct universality
classes. The cases that were discussed so far are 
summarized in Table~\ref{exptable}.

\begin{table}
\begin{tabular}{l|l|cccccc}
Situation & & $\zeta_\parallel$ & $\nu$ & $z$ & $\beta$ & $\zeta_\perp$ 
          & $z_\perp$  \cr
\hline
Anisotropic medium,    &$\kappa_{\alpha\beta}\neq0$&0.5&2&1&1&? & ?  \cr
generic direction      &              &     &    &   &    &   &    \cr
\hline
Anisotropic, FL &$\kappa_{\alpha\beta}=0$& 0.63&1.73& 1 &0.64& ?&? \cr
along symmetry axis   &$\lambda_{1\parallel}\neq0$&     &    &   &    &   
&    \cr
\hline
FL along symmetry   &$\kappa_{\alpha\beta}=0$&1&1&1.3&0.3 &0&2.3 \cr
axis, linear terms  & $\phi\neq0$ &     &    &   &    &   &    \cr
 only (Model B)&  &     &    &   &    &   &   \cr
\hline
Isotropic medium, &$\kappa_{\alpha\beta}=0$&1&1&1.3&0.3 &0.5&2.3 \cr
$\phi=0$ (Model A)&$\lambda_{1\parallel}=0$   &     &    &   &    &   &  \cr
\end{tabular}
\medskip
\caption{Critical exponents corresponding to some of the universality
classes associated with vector depinning. Entries in the first two rows
are from Ref.\protect\cite{TKD}: Transverse exponents are not known
and these cases may correspond to more than one universality class
identified by distinct $\zeta_\perp,z_\perp$.}
\label{exptable}
\end{table}

\subsection{Generalizations}
In many systems, the dynamics involves a wide range of relaxation
times. It is sometimes possible to average over ``fast" degrees of
freedom to obtain an effective equation of motion for ``slow" variables. 
For example, the motion of atoms in a metal can be described 
by an effective theory that involves only positions of the ions, assuming
that the electronic wavefunction always adjusts to
the instantaneus ionic coordinates. Similarly,
the critical dynamics of a slowly moving solid-liquid-vapor contact line 
can be described by assuming that the liquid-vapor interface
instantaneusly finds the minimum energy surface dictated by the position
of the contact line\cite{Joanny}. The elimination of these additional
degrees of freedom may cause effective nonlocal interactions between
the remaining modes, which in turn acquire a different dispersion
law. 
For example, in contact line dynamics, the elastic energy
associated with a mode of wavevector $q$ is proportional to $|q|$ 
instead of $q^2$. In general,  one may consider a situation
where the elastic energy is proportional to $|q|^\sigma$ for some 
value of $\sigma$. The scaling analysis can be easily generalized
to such cases; the most important change is the modification of
the upper critical dimension to $d_c=2\sigma$. The exponents can be easily 
calculated for general $\sigma$, as  was done by us for the 
critical dynamics of a contact line\cite{EKCL} ($\sigma=1$).

The possibility of experimental verification of our results lies in
the ability to accurately measure the motion of individual FLs
and the noise spectra (for both normal and Hall voltages) generated
by FL motion.
Very recently, there have been successful experiments that detected
the thermal motion of individual FLs at nominally zero magnetic field 
and bulk current using SQUID probes, and analyzed the noise correlation
between the two ends of the FL\cite{Lee}. A refinement of such techniques
may eventually enable a  direct comparison of theoretical results
with experiments. For example, it is known that the Hall angle changes 
sign as a function of temperature in certain 
superconductors\cite{hallswitch}. It would be particularly interesting 
to observe the increase in transverse roughness 
(thus the Hall Voltage noise) as the Hall angle approaches zero.
Ultimately, it is  very desirable to understand the properties of 
many FLs (solid or glass) near depinning, especially since this 
situation has much more experimental and technological relevance. 
One should then start from a coarse-grained
theory for the displacements ${\bf u}({\bf x}, t)$ of the 
FLs with respect to their equilibrium positions in the Abrikosov
lattice and hope to establish a similar RG scheme. However, there are 
certainly additional complications, such as 
entanglement\cite{Nelson} and plasticity\cite{Bhattacharya} effects, 
which are difficult to incorporate
in such an approach.

\acknowledgements

We have benefitted from discussions with O.~Narayan. This research was
supported by the NSF through the MRSEC Program under award number
DMR-94-00334, and via grant number DMR-93-03667.

\end{multicols}

\widetext

\begin{appendix}

\section{The Gaussian Theory}
\label{invert}

In this appendix, we compute all nonzero expectation values for
the Gaussian theory, described by the effective action
$\tilde S_0$ in Eq.(\ref{quadratic}). This is accomplished 
by inverting the quadratic form, as
\begin{eqnarray*}
\left[\matrix{\langle{\bf \hat R}({\bf q},\omega){\bf \hat R}^T 
(-{\bf q},-\omega)\rangle_c & 
\langle{\bf \hat R}({\bf q},\omega){\bf R}^T (-{\bf q},-\omega)\rangle_c  
\cr \cr
\langle{\bf R}({\bf q},\omega){\bf \hat R}^T (-{\bf q},-\omega)\rangle_c & 
\langle{\bf R}({\bf q},\omega){\bf R}^T 
(-{\bf q},-\omega)\rangle_c \cr}\right] 
&=& 
\left[\matrix{-{\bf C}(\omega) & 
{\bf G}^{-1}({\bf q},\omega)  \cr \cr
{{\bf G}^{\dag}}^{-1}({\bf q},\omega) & {\bf 0} \cr}\right]^{-1} 
\nonumber \\ \cr \nonumber \\
& &
=  \left[\matrix{ {\bf 0} & 
{\bf G}^{\dag}({\bf q},\omega)  \cr \cr
{\bf G}({\bf q},\omega) & 
{\bf G}({\bf q},\omega){\bf C}(\omega){\bf G}^{\dag}({\bf q},\omega) \cr}
\right].
\end{eqnarray*}

For the case of Model A, the individual matrices are diagonal 
and the correlation functions can be calculated easily, as
 given in Eqs.(\ref{corrA}).
For the more general case of Model B, let us first 
consider the  $v\to 0$ limit. Since $R_\perp$ occurs in the 
combination $R_\perp/v$, expectation values $\langle{\hat R}_\alpha 
R_\perp\rangle$ 
and $\langle R_\alpha R_\perp\rangle$ contribute at most $O(v)$ at the \
momentum-shell integration step. Thus, the contractions that are important 
for the momentum-shell integration 
are $\langle{\hat R}_\parallel R_\parallel\rangle$ and 
$\langle R_\parallel R_\parallel\rangle$. Setting $v=0$ and inverting the 
matrix yields

\begin{eqnarray}
{\bf G}({\bf q},\omega) &=&
\left[\matrix{K_{11}q^2-i\omega A_{11}  & 
+i\omega A_{12}  \cr 
K_{21}q^2-i\omega A_{21} & -i\omega A_{22} \cr}\right]^{-1} 
= \frac{1}{{K_\parallel}q^2-i\omega\rho_\parallel}
\left[\matrix{1  &  \kappa  \cr 
\cdots & \cdots \cr}\right], \\
{\bf G}{\bf C}{\bf G}^{\dag}({\bf q},\omega) &=&
\frac{1}{{K_\parallel}^2q^4+\omega^2\rho_\parallel^2} 
\left[\matrix{\tilde C(\omega) &  \cdots  \cr 
\cdots & \cdots \cr}\right],
\end{eqnarray}
which leads to Eqs.(\ref{corrB}).
To determine the full form of the correlation functions 
in a renormalized Gaussian theory, 
we need to perform a full matrix inversion. 
In the small $v$ limit we obtain 
\begin{eqnarray*}
\langle|R_\parallel({\bf q},\omega)|^2\rangle &=& 
\frac{1}{|\det G^{-1}|^2}\left\{
A_{22}^2\omega^2{\tilde C}(\omega)
-vq^2\omega(K_{12}A_{22}+K_{22}A_{12}){\rm Im}[C_\times(\omega)]
\right. \nonumber \\
& & \qquad \qquad \left.
+v^2q^4\left[K_{22}^2C_\parallel(\omega)-K_{22}K_{12}{\rm Re}
C_\times(\omega)+K_{12}^2C_\perp(\omega)\right]\right\},
\end{eqnarray*}
where
\begin{eqnarray*}
|\det G^{-1}|^2 &\approx& [{K_\parallel}^2q^4+\rho_\parallel^2\omega^2]
\left[v^2q^4\left([K_{11}K_{22}-K_{21}K_{12}]/{K_\parallel}\right)^2
+A_{22}^2\omega^2\right].  \\
\end{eqnarray*}
Similarly, 
\begin{eqnarray*}
\langle|R_\perp({\bf q},\omega)|^2\rangle &=&
\frac{v^2}{|\det G^{-1}|^2} \left\{
\left(K_{21}^2C_\parallel(\omega)-K_{21}K_{11}
{\rm Re}[C_\times(\omega)]
+K_{11}^2C_\perp(\omega)\right)q^4 \right.  \\
& & \qquad\qquad\qquad\qquad
-\omega q^2(K_{11}A_{21}-K_{21}A_{11}){\rm Im}[C_\times(\omega)]  \\
& &\qquad\qquad\qquad\qquad\left.+\omega^2\left(A_{21}^2C_\parallel(\omega)
-A_{21}A_{11}{\rm Re}[C_\times(\omega)]
+A_{11}^2C_\perp(\omega)\right)\right\}.
\end{eqnarray*}
At the fixed point found for Model B, Eqs.(\ref{ctilde}) are satisfied, and 
the correlation functions simplify to
\begin{eqnarray}
\langle|R_\parallel({\bf q},\omega)|^2\rangle &=&
\frac{{\tilde C}(\omega)}{{K_\parallel}^2q^4+\rho_\parallel^2\omega^2}
{\cal F}_\parallel\left({K_\perp q^2\over (\omega/v)\rho_\perp}\right), \\
\langle|R_\perp({\bf q},\omega)|^2\rangle &=&
\frac{{\tilde C}(\omega)}{4\left[
{K_\perp}^2q^4+\rho_\perp^2(\omega/v)^2\right]}
{\cal F}_\perp\left({K_\parallel q^2\over \omega\rho_\parallel}\right), 
\end{eqnarray}
where
\begin{eqnarray*}
K_\perp=&\left|
\displaystyle\frac{K_{11}K_{22}-K_{21}K_{12}}{K_{11}-\kappa K_{21}}
\right|, \qquad\qquad\qquad\qquad
\rho_\perp&=\left|
\frac{K_{11}+\kappa K_{21}}{K_{11}-\kappa K_{21}}
\right|A_{22}, \\ \\
{\cal F}_\parallel(x)=&\left[{1+x^2\left(\displaystyle
\frac{(K_{22}-K_{12}/\kappa)K_\parallel}{2(K_{11}K_{22}-K_{21}K_{12})}
\right)^2}\right]\bigg/\left[{1+x^2}\right] 
&=\cases{1,           & $x \ll 1$\cr
                               {\rm const}, & $x \gg 1$.   }  \\ \\
{\cal F}_\perp(x)    =&\left[{1+\displaystyle{1\over x^2}\left(\displaystyle
\frac{K_{11}+\kappa K_{21}}{K_{11}-\kappa K_{21}}
\frac{A_{11}-\kappa A_{21}}{A_{11}+\kappa A_{21}}
\right)^2}\right]\bigg/\left[{1+\displaystyle{1\over x^2}}\right] 
&=\cases{{\rm const}, & $x \ll 1$, \cr
                               1,           & $x \gg 1$.   }  
\end{eqnarray*}
The functions ${\cal F}_\alpha$ describe crossovers of the
overall amplitudes of the correlations,  due to the coupling
between longitudinal and transverse modes.

\section{Vertex Renormalization}
\label{Capp}

In this appendix, we derive recursion relations for the 
renormalized vertex functions $U_{\alpha,n}(u)\equiv C_{\alpha}^{(n)}(u)$. 
Let us start by considering $U_{\alpha,n}(u)$ for a given $n$.
As usual, we split the fields ${\bf R}={\bf R}^<+{\bf R}^>$ and 
${\bf \hat R}={\bf \hat R}^<+{\bf \hat R}^>$,
where fields with the superscript ``$>$" correspond to fluctuations
within the momentum shell  $\Lambda e^{-\delta\ell}<q<\Lambda$,
which are averaged over. 
In evaluating $\langle e^{\cal U}\rangle_0^>$, 
we encounter two types of nonzero contractions, 
\begin{eqnarray*}
\left<{\hat R}_\parallel^>({\bf q},t)R_\parallel^>(-{\bf q},t')\right>
&=& \frac{1}{A_{11}}\exp\left[-\frac{Kq^2(t'-t)}{A_{11}}\right]
\Theta(t'-t) \\ 
&\approx& \frac{1}{K\Lambda^2}\delta(t-t'), \\ 
\left< R_\parallel^>({\bf q},t)R_\parallel^>(-{\bf q},t')\right>
&\approx& \frac{1}{K^2\Lambda^4}U_{\alpha,0}\left(v(t-t')\right),
\end{eqnarray*}
within the momentum shell $\Lambda e^{-\delta\ell}<q<\Lambda$, and
for time scales $t-t'\sim O(1/v)$. (From now on, we suppress the subscript
0 for notational simplicity.)
Contributions to the renormalization of $U_{\alpha,n}$
come from both $\langle{\cal U}\rangle^>$ and 
$\langle{\cal U}^2\rangle_c^>$, as
\begin{eqnarray*}
\langle{\cal U}\rangle^>&=&
\sum_\alpha\frac{1}{2!(n+2)!}\int d^dx\,dt_1 dt_2 \,
U_{\alpha,n+2}(1-2)
\left<{\hat R}_\alpha(1){\hat R}_\alpha(2)[R_\parallel(1)
-R_\parallel(2)]^{n+2}
\right>^>  + \cdots \\
&=&\sum_\alpha\frac{1}{2!(n+2)!}{n+2 \choose 2}\int d^dx\,dt_1 dt_2 \,
U_{\alpha,n+2}(1-2)
{\hat R}^<_\alpha(1){\hat R}^<_\alpha(2)
\\
& & \qquad\qquad\qquad\qquad\qquad\qquad
\times[R^<_\parallel(1)-R^<_\parallel(2)]^{n}
\left<[R^>_\parallel(1)-R^>_\parallel(2)]^2\right> +\cdots, 
\end{eqnarray*}
with obvious abbreviations for the arguments of $U, R, {\hat R}$.
Evaluating the expectation values, we get
\begin{eqnarray}
\left<[R^>_\parallel(1)-R^>_\parallel(2)]^2\right> &=&
\int\limits^>{{d^d q \over (2\pi)^d}\,} 
\left\{
\left< R_\parallel^>({\bf q},t_1)R_\parallel^>(-{\bf q},t_1)\right>
+\left< R_\parallel^>({\bf q},t_2)R_\parallel^>(-{\bf q},t_2)\right>
\right. \nonumber \\
& & \qquad \qquad \qquad \qquad \left.
-2\left< R_\parallel^>({\bf q},t_1)R_\parallel^>(-{\bf q},t_2)\right>
\right\} \nonumber \\
&=& 2\delta\ell\frac{\Lambda^{d}S_d}{(2\pi)^d K^2\Lambda^4}
[U_{\parallel,0}(0)-U_{\parallel,0}(v(t_1-t_2))],
\end{eqnarray}
where $\int^>$ denotes integration over the momentum shell and 
$S_d$ is the surface area of a unit sphere in 
$d$-dimensions. Thus, the correction to  
$U^<_{\alpha,n}(u)$ from $\langle{\cal U}\rangle^>$ 
is equal to
\[
\delta\ell K_d U_{\alpha,n+2}(u)
[U_{\parallel,0}(0)-U_{\parallel,0}(u)],
\]
where $K_d\equiv\Lambda^{d-4}S_d/[(2\pi)^d K^2]$.
The contributions from $\langle{\cal U}^2\rangle_c^>$
are similarly calculated, as 
\begin{eqnarray}
\label{hugeeqn}
\langle{\cal U}^2\rangle_c^>\!\!&=&
\sum_{\alpha,\gamma}\sum_{m=1}^{n+1} \frac{1}{2!m!2!(n+2-m)!}
\int d^dx\,dt_1 dt_2 \int d^dx'\,dt_1' dt_2' \,
U_{\alpha,m}(1-2)U_{\gamma,n+2-m}(1'-2') \nonumber \\
& & \qquad\qquad \times 
\left<{\hat R}_\alpha(1){\hat R}_\alpha(2){\hat R}_\gamma(1')
{\hat R}_\gamma(2') 
[R_\parallel(1)-R_\parallel(2)]^{m}
[R_\parallel(1')-R_\parallel(2')]^{n+2-m}\right>^>+\cdots \nonumber \\
&=&\sum_{m=1}^{n+1}\frac{1}{2!(m-1)!2!(n+1-m)!}
\int d^dx\,dt_1 dt_2 \int d^dx'\,dt_1' dt_2' \,
U_{\parallel,m}(1-2)U_{\parallel,n+2-m}(1'-2') \nonumber \\
& &  \qquad\qquad \times
[R^<_\parallel(1)-R^<_\parallel(2)]^{m-1} 
[R^<_\parallel(1')-R^<_\parallel(2')]^{n+1-m} \nonumber \\
& &  \qquad\qquad \times
\left<{\hat R}_\parallel(1){\hat R}_\parallel(2){\hat R}_\parallel(1')
{\hat R}_\parallel(2')
[R^>_\parallel(1)-R^>_\parallel(2)][R^>_\parallel(1')-R^>_\parallel(2')]
\right>^>\nonumber \\
& & +2\sum_\alpha\sum_{m=1}^{n}\frac{1}{2!m!2!(n+2-m)!}{n+2-m \choose 2}
\int d^dx\,dt_1 dt_2 \int d^dx'\,dt_1' dt_2' \nonumber \\
& &  \qquad\qquad \times
U_{\alpha,m}(1-2)U_{\parallel,n+2-m}(1'-2') 
{\hat R}^<_\alpha(1'){\hat R}^<_\alpha(2')
[R^<_\parallel(1)-R^<_\parallel(2)]^{m} \nonumber \\
& &  \qquad\qquad \times
[R^<_\parallel(1')-R^<_\parallel(2')]^{n-m}
\left<{\hat R}^>_\parallel(1){\hat R}^>_\parallel(2)
[R^>_\parallel(1')-R^>_\parallel(2')]^2\right> +\cdots. 
\end{eqnarray}

The evaluations of the expectation values are tedious but
straightforward. As an example, let us evaluate the 
second half of Eq.(\ref{hugeeqn}) explicitly.
First of all,
\begin{eqnarray*}
\lefteqn{\left<{\hat R}^>_\parallel(1){\hat R}^>_\parallel(2)
[R^>_\parallel(1')-R^>_\parallel(2')]^2\right> =} \qquad & & \\
&=&
\left<{\hat R}^>_\parallel(1)R^>_\parallel(1')\right> 
\left<{\hat R}^>_\parallel(2)R^>_\parallel(1')\right> +
\left<{\hat R}^>_\parallel(1)R^>_\parallel(2')\right> 
\left<{\hat R}^>_\parallel(2)R^>_\parallel(2')\right> \\
& & \quad -2
\left<{\hat R}^>_\parallel(1)R^>_\parallel(1')\right> 
\left<{\hat R}^>_\parallel(2)R^>_\parallel(2')\right> -2
\left<{\hat R}^>_\parallel(1)R^>_\parallel(2')\right> 
\left<{\hat R}^>_\parallel(2)R^>_\parallel(2')\right>.
\end{eqnarray*}
The first two terms do not contribute to  $U^<_{\alpha,n}(u)$,
since they are proportional to $\delta(t_1-t_1')\delta(t_2-t_1')$
and $\delta(t_1-t_2')\delta(t_2-t_2')$ respectively. 
(These delta functions force $t_1$ to be equal to $t_2$. 
Since the expectation value is multiplied by  
$[R^<_\parallel(1)-R^<_\parallel(2)]^{m}$,
the final contribution is zero.) The last two terms are
equal to
\[
-2\int\limits^>{{d^d q \over (2\pi)^d}\,} \int\limits^>\frac{d^dq'}{(2\pi)^d} 
[\delta(t_1-t_1')\delta(t_2-t_2')+\delta(t_1-t_2')\delta(t_2-t_1')]
\frac{\exp\{i({\bf q}+{\bf q}')\cdot({\bf x}-{\bf x}')\}}{(Kq^2)(Kq^{'2})}.
\]
Integrating over $t_1'$, $t_2'$, ${\bf x}'$ (which yields
$\delta^d({\bf q}+{\bf q}')$) and subsequently over ${\bf q}'$,
the second half of Eq.(\ref{hugeeqn}) becomes
\begin{eqnarray*}
    -\sum_\alpha\sum_{m=1}^{n}\frac{1}{2!n!} {n \choose m} & &
\int  d^dx   dt_1 dt_2  {\hat R}^<_\alpha(1){\hat R}^<_\alpha(2)
[R^<_\parallel(1)-R^<_\parallel(2)]^{n}
\int\limits^>{{d^d q \over (2\pi)^d}\,}\frac{1}{K^2q^4} \\
& &\times\left\{
U_{\alpha,m}(1-2)U_{\parallel,n+2-m}(1-2) 
+(-1)^{n-m}U_{\alpha,m}(1-2)U_{\parallel,n+2-m}(2-1) 
\right\}
.
\end{eqnarray*}
The first half of Eq.(\ref{hugeeqn}) can be evaluated
similarly. The full contribution to $U^<_{\alpha,n}(u)$
from $\langle {\cal U}^2\rangle_c^>$  is thus equal to
\begin{eqnarray*}
-\delta\ell K_d
\left\{\delta_{\alpha,\parallel}
\sum_{m=1}^{n+1}{n \choose m-1}\right.& &(-1)^{n+2-m} 
U_{\parallel,m}(u)U_{\parallel,n+2-m}(-u)  \\
+\sum_{m=1}^{n}{n \choose m}& &\left.{1\over 2}\left[
U_{\alpha,m}(u)U_{\parallel,n+2-m}(u) +(-1)^{n+2-m}
U_{\alpha,m}(u)U_{\parallel,n+2-m}(-u)\right] \right\}.  
\end{eqnarray*}
(In the expansion of $\langle e^{\cal U}\rangle_c^>$, there 
is a factor of $1/2$ in front of 
$\langle{\cal U}^2\rangle_c^>$.) Adding all contributions,
the effective vertex function $U^<_{\parallel,n}(u)$ is 
found to be
\begin{equation}
U^<_{\parallel,n}(u)=U_{\parallel,n}(u)+
\delta\ell K_d\left\{U_{\parallel,n+2}(u)U_{\parallel,0}(0)
-\sum_{m=0}^{n+1} {n+1 \choose m} 
U_{\parallel,m}(u)U_{\parallel,n+2-m}(u) \right\},
\end{equation}
provided that
\begin{equation}
\label{Uprop}
U_{\alpha,m}(u)=(-1)^mU_{\alpha,m}(-u).
\end{equation}

Under the scale transformation (\ref{Newscale}), which brings the
momentum cutoff to its original value, we see that 
$u\to (1+\zeta_\parallel\delta\ell) u$. Thus, the renormalized vertex 
function is given by
\begin{equation}
\tilde U_{\parallel,n}(u)\equiv U_{\parallel,n}(u)+\delta\ell
\frac{\partial U_{\parallel,n}(u)}{\partial \ell}
=U^<_{\parallel,n}\left((1+\zeta_\parallel\delta\ell)u\right)
\{1+\delta\ell[d+2z_\parallel+2(\theta_\parallel-d)+n\zeta_\parallel]\}.
\end{equation}
Keeping only terms linear in $\delta\ell$, and identifying 
$U_{\parallel,n}(u)$ with the $n$th derivative of $C_\parallel(u)$,
we finally obtain the differential recursion relation for $C_\parallel(u)$:
\begin{equation}
\label{apprr}
\frac{\partial C_\parallel(u)}{\partial\ell} =
[\epsilon+2\theta_\parallel+2(z_\parallel-2)]C_\parallel(u)
+\zeta_\parallel u C'_\parallel(u) 
-K_d\left\{[C'_\parallel(u)]^2 +
[C_\parallel(u)-C_\parallel(0)]
C''_\parallel(u)\right\}.
\end{equation}
Note that the identification of $U_{\parallel,n}(u)$ with the $n$th 
derivative of $C_\parallel(u)$ is self-consistent, since recursion 
relations for $U_{\parallel,n}(u)$ are correctly recovered by taking
$n$ derivatives of Eq.(\ref{apprr}). Also, Eq.(\ref{Uprop}) is 
automatically satisfied when this identification is made since
$C_\parallel(u)=C_\parallel(-u)$.

A similar computation can be performed for $C_\perp(u)$, yielding 
\begin{equation}
U^<_{\perp,n}(u)=U_{\perp,n}(u)+\delta\ell K_d
\left\{U_{\perp,n+2}(u)U_{\parallel,0}(0)
-\sum_{m=0}^{n} {n \choose m} 
U_{\perp,n+2-m}(u)U_{\parallel,m}(u) \right\}.
\end{equation}
Upon rescaling, the renormalized vertex function is 
\begin{equation}
\tilde U_{\perp,n}(u)\equiv U_{\perp,n}(u)+\delta\ell
\frac{\partial U_{\perp,0}(u)}{\partial \ell}
=U^<_{\perp,n}\left((1+\zeta_\parallel\delta\ell)u\right)
\{1+\delta\ell[d+2z_\parallel+2(\theta_\perp-d)+n\zeta_\parallel]\}.
\end{equation}
Thus, we obtain the recursion relation
\begin{equation}
\frac{\partial C_\perp(u)}{\partial\ell} =
[\epsilon+2\theta_\perp+2(z_\parallel-2)]C_\perp(u)
+\zeta_\parallel u C'_\perp(u) 
-K_d\left\{[C_\parallel(u)-C_\parallel(0)]
C''_\perp(u)\right\}.
\end{equation}

\section{Higher-Order Diagrams}
\label{Cperpapp}
In this appendix, we show that the sum of all  
contributions to the renormalization of $C_\perp(u)$ from the 
momentum shell integration step vanish in the limit 
$u\to 0^+$. This was already explicitly demonstrated for the 
leading order contributions that come from $\langle{\cal U}\rangle_c$
and $\langle{\cal U}^2\rangle_c$. Since the only nonzero contractions 
involve $R_\parallel$ and ${\hat R}_\parallel$, 
all contributions to the renormalization of $C_\perp(v(t-t'))$ due to
$\left<{\cal U}^m\right>_c$ arise from terms of the form
\begin{eqnarray*}
\left<{\cal U}^m\right>_c&=&\sum_{n=2}^{\infty}\int {d^d\!x\,}{dt\,} dt'
\frac{U_{\perp,n}(v(t-t'))}{2!n!} {\hat R}^<_\perp(t){\hat R}^<_\perp(t')
\int \left[
\prod_{i=1}^{m-1}d^d\!x_i\, dt_i\, dt_i'\,
\frac{ U_{\parallel,n_i}(v(t_i-t_i')) }{2!n_i!}
\right]\\
& & \qquad  \times
\left<[R_\parallel^>({\bf x}, t)-R_\parallel^>({\bf x}, t')]^n 
\prod_{i=1}^{m-1}{\hat R}_\parallel^>({\bf x}_i, t_i)
{\hat R}_\parallel^>({\bf x}_i, t'_i)
[R_\parallel^>({\bf x}_i, t_i)-R_\parallel^>({\bf x}_i, t'_i)]^{n_i}
\right>_c+\cdots.
\end{eqnarray*}

\begin{multicols}{2}

The expectation value clearly goes to zero as $(t-t')^n$ in the 
$t\to t'^+$ limit. This gives us the desired result that 
$C_\perp(0)$  is unrenormalized to all orders in perturbation theory.

\section{High-frequency structure of $U_{\parallel,2}$}
\label{hifreq}
In this appendix, we shall demonstrate that there are no $v^{-1}$
divergences in the renormalization of $A_{11}$, at least to
$O(\epsilon)$. In order to do this, we examine the full form
of the bare vertex function $U_{\parallel,2}$ obtained from 
MF theory,
\[
U_{\parallel,2}(t_1,t_2;t'_1,t'_2)=\frac{\partial^2
\left<\overline r_\parallel(t_1)
\overline r_\parallel(t_2)\right>_{MF,c}}
{\partial\varepsilon_\parallel(t'_1)\partial\varepsilon_\parallel(t'_2)} .
\]
The low-frequency analysis of this vertex function gives a result
proportional to $1/v$ when all times are within $O(1)$ of each
other. This may potentially give an $O(1/v)$ contribution to the
renormalization of $A_{11}$. Indeed, an external impulse of magnitude 
$\varepsilon$ right before a ``jump" (the fast motion between
consequent local minima) shifts 
the jump time by $\varepsilon/\eta$ and creates a response of
$O(1/v)$ right after the jump takes place. However, an impulse 
right after a jump does not affect the jump time and creates a 
response of only $O(1)$. Thus a singular response is seen
if all times are in the vicinity of a jump, say, at time $t_J$. 
$U_{\parallel,2}(t_1,t_2;t'_1,t'_2)$ can be as large 
as $O(v^{-2})$ if $t'_1$ and $t'_2$ are both slightly less than 
$t_J$, and $t_1$ and $t_2$ are both slightly greater than $t_J$.
Considering that the probability of being close to a jump is $v$, 
this term can potentially contribute as much as $O(v^{-1})$ to the 
renormalization of $A_{11}$ upon statistical time-averaging.
A careful analysis, and explicit evaluation of this vertex in 
the case of a periodic potential\cite{NFAppendix}, show that this
is the only way a singularity may occur in the RG
contributions. However, when the times $t'_1,t'_2$ of fields 
$R_\parallel$ are smaller than the times $t_1,t_2$ of fields 
${\hat R}_\parallel$, the contraction 
$\left<{\hat R}_\parallel(t_i)R_\parallel(t'_j)\right>_0$ which appears
in the RG contribution is identically zero due to the
causality of the propagator. 
Therefore, the singular part of $C_\parallel^{*''}(0)$ 
does not enter the renormalization of $A_{11}$ 
(or $\rho_\parallel$ in the case of Model B) to one-loop order. 

\section{Renormalization of Model B}
\label{modelB}

Details of the RG calculation for Model B are presented in this
appendix. For the sake of brevity, we shall only consider the 
renormalization
of the parameters in the Gaussian theory, i.e. the propagator,
and the two-point correlation functions 
$U_{\parallel,0}(u)$, $U_{\perp,0}(u)$, $U_{\times,0}(u)$. The 
renormalization
of higher-order vertex functions are again related to derivatives of
$C_\alpha$ through
$U_{\alpha,n}(u)\equiv C_\alpha^{(n)}(u)$.

Nonzero contractions involved in the calculation are given 
in Eqs.(\ref{corrB}). The parameters $A_{12},A_{22}$ (thus $\kappa$), and
$K_{\alpha\gamma}$ (thus $K_\parallel, K_\perp$, and $\rho_\perp$) 
do not get contributions from the momentum shell
integration, and give rise to exponent identities discussed in 
the text. On the other hand, $A_{11}$ and $A_{21}$ (thus $\rho_\parallel$), 
as well as the functions $C_\alpha(u)$, are renormalized. 
Let us start by looking at the renormalization of two-point correlation 
functions $C_\alpha(u)$. By definition, $C_\parallel(u)=C_\parallel(-u)$ 
and $C_\perp(u)=C_\perp(-u)$, but $C_\times(u)\neq C_\times(-u)$ in general. 
It is convenient to write $C_\times(u)$ 
in terms of its even and odd parts $C_{\times S}(u)$ and $C_{\times A}(u)$ 
respectively, and follow their renormalization separately. 

The momentum shell integration procedure is similar to the one 
presented in Appendix \ref{Capp}, albeit more cumbersome due to 
many more nonzero contractions. Nevertheless, 
carrying out the computation yields
\begin{equation}
C^<_\alpha(u)=C_\alpha(u)-\delta\ell K_d {\cal I}_\alpha(u), 
\end{equation}
for $u>0$, where
\end{multicols}

\begin{eqnarray}
{\cal I}_\parallel(u) &=&
C_\parallel''(u)[{\tilde C}(u)-{\tilde C}(0^+)]
+C_\parallel'(u){\tilde C}'(u)\nonumber \\
& & \quad -\kappa^2\left\{C_\parallel'(u)C_\perp'(u)-[C_{\times S}'(u)/2]^2
+[C_{\times A}'(u)/2]^2 \right\} \nonumber \\
& & \quad
+\kappa C_{\times A}(u)
\left\{C_\parallel'(0^+)+\kappa[C_{\times A}'(0^+)/2]\right\}, 
\\
{\cal I}_\perp(u) &=&
C_\perp''(u)[{\tilde C}(u)-{\tilde C}(0^+)]
+C_\perp'(u){\tilde C}'(u)\nonumber  \\
& & \quad
-\left\{C_\parallel'(u)C_\perp'(u)-[C_{\times S}'(u)/2]^2
+[C_{\times A}'(u)/2]^2 \right\} \nonumber \\
& & \quad
+C_{\times A}(u)
\left\{[C_{\times A}'(0^+)/2]-\kappa C_\perp'(0^+)\right\}, 
\\
{\cal I}_{\times S}(u) &=&
C_{\times S}''(u)[{\tilde C}(u)-{\tilde C}(0^+)]
+C_{\times S}'(u){\tilde C}'(u)\nonumber \\
& & \quad
+2\kappa\left\{C_\parallel'(u)C_\perp'(u)-[C_{\times S}'(u)/2]^2
+[C_{\times A}'(u)/2]^2 \right\}\nonumber \\
& & \quad
-C_{\times A}(u)
\left\{C_\parallel'(0^+) +\kappa C_{\times A}'(0^+) -\kappa^2 
C_\perp'(0^+)\right\}, 
\\
{\cal I}_{\times A}(u) &=&
C_{\times A}''(u)[{\tilde C}(u)-{\tilde C}(0^+)]
+C_{\times A}'(u){\tilde C}'(u)\nonumber \\
& & \quad
+4\kappa\left\{C_\parallel'(u)C_\perp'(0^+)
-C_\parallel'(0^+)C_\perp'(u)\right\}\nonumber \\
& & \quad
-C_{\times S}'(u)\left\{C_\parallel'(0^+)+\kappa C_{\times A}'(u)
-\kappa^2C_\perp'(0^+)\right\}.
\end{eqnarray}
Thus, the renormalization of ${\tilde C}(u)$ is given by
\begin{equation}
{\tilde C}^<(u)={\tilde C}(u)-\delta\ell K_d 
\left\{{\tilde C}''(u)[{\tilde C}(u)-{\tilde C}(0^+)]
+[{\tilde C}'(u)]^2\right\}, 
\end{equation}
which leads to the functional recursion relation
\begin{equation}
\frac{\partial {\tilde C}(u)}{\partial\ell} =
[\epsilon+2\theta_\parallel+2(z_\parallel-2)]{\tilde C}(u)
+\zeta_\parallel u {\tilde C}'(u) 
-K_d\left\{[{\tilde C}'(u)]^2 +
[{\tilde C}(u)-{\tilde C}(0)]
{\tilde C}''(u)\right\}.
\end{equation}
This is identical to Eq.(\ref{apprr}), with the substitution 
$C_\parallel(u)\to{\tilde C}(u)$. It is straightforward to verify 
that there exists a fixed point where individual matrix elements 
$C_\alpha(u)$ satisfy Eq.(\ref{ctilde}). 
($C_{\times A}(u)=0$ at this fixed point.)

Let us next examine the renormalization of $\rho_\parallel$.
Leading order contributions come from $\langle {\cal U}\rangle^>_0$,
and a calculation along the lines presented in Sec.\ref{scaling}
gives
\begin{eqnarray*}
A_{11}^< &=& A_{11}-\delta\ell \frac{S_d\Lambda^{d}}
{(2\pi)^{d}\rho_\parallel}
\int\limits_0^\infty d\tilde t\,\tilde t\,
e^{-K_\parallel\Lambda^2\tilde t/\rho_\parallel}
\left[C''_\parallel(v\tilde t\,)
+{\kappa\over 2}C''_\times(v\tilde t\,)\right], \\
A_{21}^< &=& A_{21}-\delta\ell \frac{S_d\Lambda^{d}}
{(2\pi)^{d}\rho_\parallel}
\int\limits_0^\infty d\tilde t\,\tilde t\,
e^{-K_\parallel\Lambda^2\tilde t/\rho_\parallel}
\left[{1\over 2}C''_\times(-v\tilde t\,)
+\kappa C''_\perp(v\tilde t\,)\right], 
\end{eqnarray*} 
which can be combined to yield
\begin{equation}
\rho_\parallel^< = \rho_\parallel-\delta\ell 
\frac{S_d\Lambda^{d}}{(2\pi)^{d}\rho_\parallel}
\int\limits_0^\infty d\tilde t\,\tilde t\,
e^{-K_\parallel\Lambda^2\tilde t/\rho_\parallel}
{\tilde C}''(v\tilde t\,). 
\end{equation}
The fixed-point function ${\tilde C}^*(u)$ is identical to
that of $C^*_\parallel(u)$ in Model A, and its behavior near
$u=0$ is also given by Eq.(\ref{Corigin}).  
Thus, we obtain 
\[
\rho_\parallel^< = \rho_\parallel
-\delta\ell \rho_\parallel K_d {\tilde C}''(0^+)
= \rho_\parallel[1-\delta\ell(2\epsilon/9)]
\]
which leads to the recursion relation
\begin{equation}
\frac{\partial \rho_\parallel}{\partial \ell}=
\rho_\parallel[\theta_\parallel+\zeta_\parallel-2\epsilon/9].
\end{equation}
\end{appendix}

\begin{multicols}{2}

\end{multicols}


\begin{references}
\bibitem{backwards}This is not strictly true, but backward motion happens
very rarely and can be ignored at larger length scales, since the 
probability of backward motion decreases exponentially at increasing 
length scales.
\end{references}
\end{document}